\documentclass[twocolumn,showpacs,preprintnumbers,amsmath,amssymb]{revtex4}
\usepackage{graphicx}% Include figure files
\usepackage{dcolumn}% Align table columns on decimal point
\usepackage{bm}% bold math

\begin{document}
%\preprint{APS/123-QED}
\title{Theory of light-induced current in molecular-tunneling junctions excited with
intense shaped pulses}
\author{B. D. Fainberg }
\affiliation{Faculty of Sciences, Holon Institute of Technology, 52 Golomb St., Holon
58102, Israel }
\affiliation{Raymond and Beverly Sackler Faculty of Exact Sciences, School of Chemistry,
Tel-Aviv University, Tel-Aviv 69978, Israel}
\author{M. Jouravlev and A. Nitzan}
\affiliation{Raymond and Beverly Sackler Faculty of Exact Sciences, School of Chemistry,
Tel-Aviv University, Tel-Aviv 69978, Israel}

\begin{abstract}
A theory for light-induced current by strong optical pulses in
molecular-tunneling junctions is described. We consider a molecular bridge
represented by its highest occupied and lowest unoccupied levels, HOMO and
LUMO, respectively. We take into account two types of couplings between the
molecule and the metal leads: electron transfer that gives rise to net current
in the biased junction and energy transfer between the molecule and
electron-hole excitations in the leads. Using a Markovian approximation, we
derive a closed system of equations for the expectation values of the relevant
variables: populations and molecular polarization that are binary, and exciton
populations that are tetradic in the annihilation and creation operators for
electrons in the molecular states. We have proposed an optical control method
using chirped pulses for enhancing charge transfer in unbiased junctions where
the bridging molecule is characterized by a strong charge-transfer transition.
An approximate analytical solution of the resulting dynamical equation is
supported by a full numerical solution. When energy transfer between the
molecule and electron-hole excitations in the leads is absent, the optical
control problem for inducing charge transfer with linearly chirped pulse can
be reduced to the Landau-Zener transition to a decaying level. When chirp is
fast with respect to the rate of the electron transfer, the Landau theory is
recovered. The proposed control mechanism is potentially useful for developing
novel opto-electronic single-electron devices with optical gating based on
molecular nanojunctions.

\end{abstract}

\maketitle

\section{Introduction}

Molecular electronics is one of the most promising substitutes for today's
semiconductor electronics. In this relation molecular conduction nanojunctions
have been under intense study in the last few years
\cite{Kagan_Ratner04,Joachim_Ratner05,Hanggi02,Hanggi04}. Recently, light
induced switching behavior in the conduction properties of molecular
nanojunctions has been demonstrated
\cite{Dulic03,Wakayama04,Yasutomi04,Wu06,Katsonis06,He05}.

However, the use of an external electromagnetic field as a controlling tool in
the small nanogap between two metal leads is difficult to implement. Currently
techniques available to achieve high spatial resolution with laser
illumination are limited by diffraction to about half of the optical
wavelength. The introduction of near-field scanning optical microscopes (NSOM)
and tip-enhanced NSOM \cite{Hartschuh04} has extended the spatial resolution
beyond the diffraction limit. The latter technique uses the strongly confined
electromagnetic field generated by optically exciting surface plasmons
localized at the apex of a sharp metallic tip, increasing spatial resolution
to better than 10 nm \cite{Hartschuh04}. Recently, spatial resolution at the
atomic scale has been also achieved in the coupling of light to single
molecules adsorbed on a surface, using scanning tunneling microscopy (STM)
\cite{Wu06}.

If experimental setups that can couple biased molecular wires to the radiation
field could be achieved, general questions concerning current through the
molecular nanojunctions in nonequilibrium situations come to mind. Recently
Galperin and Nitzan investigated a class of molecules characterized by strong
charge-transfer transitions into their first excited state \cite{Nit05}. The
dipole moment of such molecules changes considerably upon excitation,
expressing a strong shift of the electronic charge distribution. For example,
the dipole moment of 4-dimethylamino-4'-nitrostilbene (DMEANS) is 7 D in the
ground state and $\sim$ 31 D in the first excited singlet state
\cite{Smirnov98}. For all-trans retinal in polymethyl methacrylate films the
dipole increases from $\sim$6.6 to 19.8 D upon excitation to the $^{1}B_{u}$
electronic state \cite{Pondert83} and 40 {\AA } CdSe nanocrystals change their
dipole from $\sim$0 to $\sim$32 D upon excitation to their first excited state
\cite{Colvin92}. In the independent electron picture this implies that either
the highest occupied, or the lowest unoccupied, molecular orbitals (HOMO,
$|1\rangle$, or LUMO, $|2\rangle$, see Fig.\ref{fig:model}) is dominated by
atomic orbitals of larger amplitude (and better overlap with metal orbitals)
on one side of the molecule than on the other and therefore, when used as
molecular wires connecting two metal leads, stronger coupling to one of the
leads. They have shown that when such molecular wire connects between two
metal leads, weak steady-state optical pumping can create an internal driving
force for charge flow between the leads.

A theory of light-induced effects by weak CW radiation in molecular
conduction was developed in Ref.\cite{Nitzan06JCP}. However, there
are reasons to consider also molecular junctions subjected to strong
electromagnetic fields. First, the structure of such junctions is
compatible with configurations considered for large electromagnetic
field as in tip enhanced NSOM \cite{Hartschuh04}. Secondly, it was
demonstrated in Ref.\cite{Hartschuh04} that the combination of
near-field optics and ultrafast spectroscopy is readily achieved,
and the observation of photo-induced processes, such as charge
transfer, energy transfer or isomerization reactions on the
nanoscale is feasible \cite{Brixner06}. Third, consideration of
junction stability and integrity suggests that strong radiation
fields should be applied as sequences of well separated pulses to
allow for sufficient relaxation and heat dissipation. Finally,
consideration of strong time dependent pulses makes it possible to
study ways to optimize the desired effect, here the light induced
electron tunneling, i.e. to explore possibilities of coherent
control of charge flow between the leads. Our objective in the
present work is to extend the theory of
Refs.\cite{Nit05,Nitzan06JCP} to strong fields and to apply the
theory to studies of coherent control of nanojunction transport.

While these problems are of general and fundamental interest, we note that
this study is related to efforts to develop novel optoelectronic
single-electron devices, such as a photon--electron conversion device, optical
memory, and single-electron transistors with optical gating \cite{Wakayama04}.
In addition, the potential significance of molecular nanojunctions for device
applications lies in the possibility of creating all-optical switches
\cite{Hanggi03} that could be incorporated in future generations of optical
communications systems. It is conceivable that these devices will employ
coherent optical manipulations, because the speed of coherent manipulations
greatly exceeds that of currently available electronic devices.

The outline of the paper is as follows. In Sec.\ref{sec:Hamiltonian} we
introduce our model. In Sec.\ref{sec:equations} we derive a closed set of
equations for the expectation values of binary and tetradic variables of the
annihilation and creation operators for electrons in molecular states
$|1\rangle$ and $|2\rangle,$ and get formulas for the current and charge
transferred during the electromagnetic pulse action. In
Sec.\ref{sec:Strong_field} we calculate a current induced by quasistationary
intense light pulse. Optical control of current and transferred charge with
chirped pulses is considered in Sec.\ref{sec:Optical control}. We summarize
our results in Sec.\ref{sec:Conclusion}. In the Appendices we show that in the
absence of the radiative and nonradiative energy transfer couplings, the
equations of motion derived in the paper lead to the well known Landauer
formula for the current and present auxiliary calculations.

\section{The model Hamiltonian}

\label{sec:Hamiltonian}

\begin{figure}
[ptb]
\begin{center}
\includegraphics[width=3.4in]%
{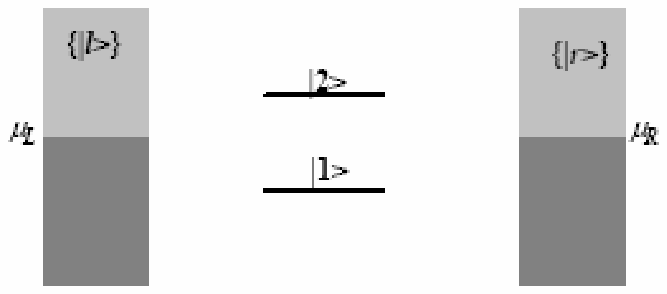}%
\caption{ A model for light induced effects in molecular conduction. The right
($R=|\{r\}\rangle$ ) and left ($L=|\{l\}\rangle$) manifolds represent the two
metal leads characterized by electrochemical potentials $\mu_{R}$ and $\mu
_{L}$ respectively. The molecule is represented by its highest occupied
molecular orbital (HOMO), $|1\rangle$, and lowest unoccupied molecular orbital
(LUMO), $|2\rangle$.}%
\label{fig:model}%
\end{center}
\end{figure}

We consider a system that comprises a molecule represented by its highest
occupied molecular orbital (HOMO), $|1\rangle$ , and lowest unoccupied
molecular orbital (LUMO), $|2\rangle$ , positioned between two leads
represented by free electron reservoirs $L$ and $R$ and interacting with the
radiation field (Fig.\ref{fig:model}). In the independent electron picture a
transition between the ground and excited molecular states corresponds to
transfer of an electron between levels $|1\rangle$ and $|2\rangle$. The
electron reservoirs (leads) are characterized by their electronic chemical
potentials $\mu_{L}$ and $\mu_{R}$, where the difference $\mu_{L}-\mu_{R}$
$=e\Phi$ is the imposed voltage bias.%
%TCIMACRO{\FRAME{ftbpFU}{3.0902in}{1.8795in}{0pt}{\Qcb{ A model for light
%induced effects in molecular conduction. The right ($R=|\{r\}\rangle$ ) and
%left ($L=|\{l\}\rangle$) manifolds represent the two metal leads characterized
%by electrochemical potentials $\mu_{R}$ and $\mu_{L}$ respectively. The
%molecule is represented by its highest occupied molecular orbital (HOMO),
%$|1\rangle$, and lowest unoccupied molecular orbital (LUMO), $|2\rangle$.}%
%}{\Qlb{fig:model}}{fig_model.eps}{\special{ language "Scientific Word";
%type "GRAPHIC";  maintain-aspect-ratio TRUE;  display "USEDEF";
%valid_file "F";  width 3.0902in;  height 1.8795in;  depth 0pt;
%original-width 3.1204in;  original-height 1.8866in;  cropleft "0";
%filename 'Fig_model.eps';file-properties "NPEU";}}}%
%BeginExpansion
%EndExpansion
The Hamiltonian is%

\begin{equation}
\hat{H}=\hat{H}_{0}+\hat{V} \label{eq:H}%
\end{equation}
where%

\begin{equation}
\hat{H}_{0}=\sum_{m=1,2}\varepsilon_{m}\hat{n}_{m}+\sum_{k\in\{L,R\}}%
\varepsilon_{k}\hat{n}_{k} \label{eq:Ho}%
\end{equation}
contains additively terms that correspond to the isolated molecule ($m$) and
the free leads ($k$). Here $\hat{n}_{i}=\hat{c}_{i}^{+}\hat{c}_{i}$ is the
population operator in state $i$, the operators $\hat{c}$ and $\hat{c}^{+}$
are annihilation and creation operators of an electron in the various states.

The interaction term $\hat{V}$ can be written as%
\begin{equation}
\hat{V}=\hat{V}_{M}+\hat{V}_{P}+\hat{V}_{N} \label{eq:V}%
\end{equation}
where $\hat{V}_{P}$ accounts for the effect of the external radiation field.
The latter is represented by the (classical) function%
\begin{equation}
\mathbf{E}(\mathbf{r},t)=\mathbf{E}^{(+)}(t)+\mathbf{E}^{(-)}(t)=\frac{1}%
{2}\mathbf{e}\mathcal{E}(t)\exp[-i\omega t+i\varphi(t)]+c.c.
\label{eq:EmField1}%
\end{equation}
characterized by the pulse envelope $\mathcal{E}(t)$, carrier frequency
$\omega$ and (possibly) time dependent phase $\varphi(t)$. The time dependent
phase corresponds to time evolution of the pulse frequency (chirp)
$\omega(t)=\omega-d\varphi(t)/dt$. Introducing bilinear operators of the
excitonic type%

\begin{equation}
b_{M}^{+}=\hat{c}_{2}^{+}\hat{c}_{1},\text{ }b_{M}=\hat{c}_{1}^{+}\hat{c}%
_{2},\text{ }b_{ij}^{+}=\hat{c}_{i}^{+}\hat{c}_{j}=b_{ji}\text{, \ (}i\neq
j\text{),} \label{eq:exciton}%
\end{equation}
the molecule-radiation field coupling, $\hat{V}_{P}$, can be written as
follows in the resonance or rotating wave approximation (RWA)%
\begin{equation}
\hat{V}_{P}=-\frac{1}{2}(\mathbf{d}\cdot\mathbf{e)}\{b_{M}^{+}\mathcal{E}%
(t)\exp[-i\omega t+i\varphi(t)]+h.c.\} \label{eq:V_P_res}%
\end{equation}
where $\mathbf{d}$ is the transition dipole moment.

The other terms in Eq.(\ref{eq:V}) describe coupling between the molecule and
the metal electronic subsystems. In terms of the excitonic operators defined
in Eq.(\ref{eq:exciton}), they are given by%

\begin{equation}
\hat{V}_{M}=\sum_{K=L,R}\sum_{m=1,2;k\in K}(V_{km}^{(MK)}b_{mk}+h.c.)\text{,}
\label{eq:V_M}%
\end{equation}

\begin{equation}
\hat{V}_{N}=\sum_{K=L,R}\sum_{k\neq k^{\prime}\in K}(V_{kk^{\prime}}%
^{(NK)}b_{k^{\prime}k}b_{M}^{+}+V_{k^{\prime}k}^{(NK)}b_{M}b_{k^{\prime}k}%
^{+})\text{,} \label{eq:V_N_exciton1}%
\end{equation}
$L$ and $R$ denote the left and right leads, respectively, and $h.c.$ denotes
Hermitian conjugate. $\hat{V}_{M}$ and $\hat{V}_{N}$, Eqs. (\ref{eq:V_M}) and
(\ref{eq:V_N_exciton1}), respectively, denote two types of couplings between
the molecule and the metal leads: $\hat{V}_{M}$ describes electron transfer
that gives rise to net current in the biased junction, while $\hat{V}_{N}$
describes energy transfer between the molecule and electron-hole excitations
in the leads. The latter interaction strongly affects the lifetime of excited
molecules near metal surfaces. $\hat{V}_{N}$ is written in the near field
approximation, disregarding retardation effects that will be important at
large molecule-lead distances.

\section{Equations of motion}

\label{sec:equations}

The physics of the system can be described within different approaches. One is
the method of nonequilibrium Green's functions
\cite{Nit05,Nitzan06JCP,Harbola_Mukamel06}. It has advantages of a formal
treatment due to the possibility of a diagrammatic representation, and it is
particularly well suited for stationary processes where the Dyson equation can
be cast in the energy representation. For time-dependent processes, such as
are the subject of this work, a method based on the equations of motion for
the expectation values of the operators provides a more transparent approach,
since the quantities are more directly related to physical observables. Such a
method is adopted here. Using a Markovian approximation for the relaxation
induced by the molecule-metal leads coupling, we derive a closed set of
equations for the expectation values of binary $\langle\hat{n}_{m}%
\rangle=n_{m}$ and $\langle b_{M}\rangle=p_{M}$, and tetradic $\langle
b_{M}^{+}b_{M}\rangle=N_{M}$ variables of the annihilation and creation
operators for electrons in molecular states $|1\rangle$ and $|2\rangle$. The
first expression is simply the population of electrons in molecular state $m$,
the second gives the molecular polarization and the third represents the
molecular excitation, referred to below as the molecular exciton population.

Using the Heisenberg equations of motion one obtains the equation for the
expectation value of any operator $\hat{F}$%
\begin{equation}
\frac{d}{dt}\langle\hat{F}\rangle=\frac{i}{\hbar}\langle\lbrack\hat{H}%
_{0}+\hat{V},\hat{F}]\rangle\equiv\frac{i}{\hbar}Tr([\hat{H}_{0}+\hat{V}%
,\hat{F}]\rho) \label{eq:Heis1}%
\end{equation}
where $\rho$ is the density matrix. Straightforward operator algebra
manipulations yield for $n_{m}$ and $p_{M}$ in RWA%

\begin{align}
\frac{dn_{m}}{dt}  &  =(-1)^{m}\operatorname{Im}\{\Omega^{\ast}(t)p_{M}%
\exp[i\omega t-i\varphi(t)]\}-\nonumber\\
&  -\frac{2}{\hbar}\operatorname{Im}\sum_{K=L,R}\sum_{k\in K}V_{km}%
^{(MK)}\langle b_{mk}\rangle-\nonumber\\
&  -\frac{2}{\hbar}\operatorname{Im}\sum_{K=L,R}\sum_{k\neq k^{\prime}\in
K}[\delta_{2m}V_{k^{\prime}k}^{(NK)}\langle b_{M}b_{k^{\prime}k}^{+}%
\rangle+\nonumber\\
&  +\delta_{1m}V_{kk^{\prime}}^{(NK)}\langle b_{M}b_{k^{\prime}k}^{+}%
\rangle^{\ast}] \label{eq:n_m}%
\end{align}
\begin{widetext}
\begin{eqnarray}
\frac{dp_{M}}{dt}  &  =\frac{i}{\hbar}(\varepsilon_{1}-\varepsilon_{2}%
)p_{M}+\frac{i}{2}\Omega(t)\exp[-i\omega t+i\varphi(t)](n_{1}-n_{2}%
)+\frac{i}{\hbar}\sum_{K=L,R}\sum_{k\in K}(V_{k1}^{(MK)}\langle b_{k2}%
^{+}\rangle-V_{2k}^{(MK)}\langle b_{k1}\rangle)+\nonumber\\
&  +\frac{i}{\hbar}\sum_{K=L,R}\sum_{k\neq k^{\prime}\in K}V_{kk^{\prime}%
}^{(NK)}\langle b_{k^{\prime}k}(\hat{n}_{2}-\hat{n}_{1})\rangle\label{eq:p_M4}%
\end{eqnarray}
\end{widetext}
where $\Omega(t)=(\mathbf{d}\cdot\mathbf{e)}\mathcal{E}(t)/\hbar$ is the Rabi
frequency. The equations of motion include couplings to additional
correlations of the second order $\langle b_{mk}\rangle$ due to the electron-
transfer interaction $\hat{V}_{M}$, and to higher-order correlations $\langle
b_{M}b_{k^{\prime}k}^{+}\rangle$ etc. due to the energy transfer $\hat{V}%
_{N}.$ To obtain expressions for these correlations, we now compute their
equations of motion, using the Markovian approximation for the relaxations
induced by the molecule-metal leads couplings, $\hat{V}_{M}$ and $\hat{V}_{N}%
$. In this work we assume that the relaxation processes due to $\hat{V}_{M}$
and $\hat{V}_{N}$ are not interdependent and also do not depend on the
external electromagnetic field. We shall discuss the last approach in
Sec.\ref{sec:Conclusion}

\subsection{Calculation of terms associated with the electron transfer
interaction $\hat{V}_{M}$ in the equations for $n_{m}$ and $p_{M}$}

\label{subsec:Mkrelaxation_for_n_and_p}

In evaluating the effect of the relaxation processes associated with the
electron transfer and energy transfer terms in the Hamiltonian, $\hat{V}_{M}$
and $\hat{V}_{N}$ respectively, we make the approximation (known as the
non-crossing approximation) that these processes do not affect each other. A
similar assumption is made with respect to the effect of the external field.
With this in mind we consider the expectation values $\langle b_{km}\rangle$
and $\langle b_{mk}\rangle$ that enter the terms containing $\hat{V}_{M}$ on
the right-hand side of Eqs.(\ref{eq:n_m}) and (\ref{eq:p_M4}) and omit
$\hat{V}_{P}$ and $\hat{V}_{N}$ terms in the equations of motion that describe
their evolution. This leads to%
\begin{align}
\frac{d}{dt}\langle b_{mk}\rangle &  =\frac{i}{\hbar}(\varepsilon
_{k}-\varepsilon_{m})\langle b_{mk}\rangle+\frac{i}{\hbar}\sum_{m^{\prime
}=1,2}V_{m^{\prime}k}^{(MK)}\langle\hat{c}_{m^{\prime}}^{+}\hat{c}_{m}%
\rangle-\nonumber\\
&  -\frac{i}{\hbar}V_{mk}^{(MK)}f_{K}(\varepsilon_{k}) \label{eq:c^+_kc_m}%
\end{align}
where we assumed that the leads are in equilibrium with the expectation values%

\begin{equation}
\langle\hat{c}_{k}^{+}\hat{c}_{k^{\prime}}\rangle=f_{K}(\varepsilon_{k}%
)\delta_{kk^{\prime}}, \label{eq:c^+_kc_k'}%
\end{equation}

\[
f_{K}(\varepsilon_{k})=[\exp((\varepsilon_{k}-\mu_{K})/k_{B}T)+1]^{-1}%
\]
is the Fermi function, $\delta_{kk^{\prime}}$ is the Kronecker delta. Formally
integrating Eq.(\ref{eq:c^+_kc_m}), we get%

\begin{align}
\langle b_{mk}\rangle &  =\frac{i}{\hbar}\int_{-\infty}^{t}dt^{\prime}%
\exp[\frac{i}{\hbar}(\varepsilon_{k}-\varepsilon_{m})(t-t^{\prime}%
)]\times\nonumber\\
&  \times\lbrack\sum_{m^{\prime}=1,2}V_{m^{\prime}k}^{(MK)}\langle\hat
{c}_{m^{\prime}}^{+}\hat{c}_{m}\rangle(t^{\prime})-V_{mk}^{(MK)}%
f_{K}(\varepsilon_{k})] \label{eq:b_mk}%
\end{align}
In the absence of the $\hat{V}_{M}$ coupling this results in a set of
integro-differential equations for $\langle\hat{c}_{m}^{+}\hat{c}_{m}%
\rangle=n_{m}$ and $\langle\hat{c}_{1}^{+}\hat{c}_{2}\rangle=p_{M}$. The
dynamics contains memory effects, and therefore is non-Markovian. Next we make
a Markovian approximation by transforming $\langle\hat{c}_{m^{\prime}}^{+}%
\hat{c}_{m}\rangle$ to the interaction representation: $\langle\hat
{c}_{m^{\prime}}^{+}\hat{c}_{m}\rangle(t^{\prime})=\langle\hat{c}_{m^{\prime}%
}^{+}\hat{c}_{m}\rangle^{int}(t^{\prime})\exp[\frac{i}{\hbar}(\varepsilon
_{m^{\prime}}-\varepsilon_{m})t^{\prime}]$ and assuming that slowly varying
function $\langle\hat{c}_{m^{\prime}}^{+}\hat{c}_{m}\rangle^{int}(t^{\prime})$
can be moved as $\langle\hat{c}_{m^{\prime}}^{+}\hat{c}_{m}\rangle^{int}(t)$
to outside the integral \footnote{Strictly speaking, this approximation is
valid only for the sums $\sum_{k}$ that appear in
Eqs.(\ref{eq:n_m_MKrelaxation}) and (\ref{eq:p_MKrelaxation}) (provided that
the manifold $\{k\}$ constitutes a smooth continuum). Indeed
Eq.(\ref{eq:b_mk1}) is meaningful only within such sums.}. Eq.(\ref{eq:b_mk})
then becomes%
\begin{align}
\langle b_{mk}\rangle &  =\langle b_{km}\rangle^{\ast}=\frac{i}{\hbar}%
\{\sum_{m^{\prime}=1,2}V_{m^{\prime}k}^{(MK)}\langle\hat{c}_{m^{\prime}}%
^{+}\hat{c}_{m}\rangle(t)\times\nonumber\\
&  \times\lbrack\frac{i\hbar P}{\varepsilon_{k}-\varepsilon_{m^{\prime}}%
}+\hbar\pi\delta(\varepsilon_{k}-\varepsilon_{m^{\prime}})]-\nonumber\\
&  -V_{mk}^{(MK)}f_{K}(\varepsilon_{k})[\frac{i\hbar P}{\varepsilon
_{k}-\varepsilon_{m}}+\hbar\pi\delta(\varepsilon_{k}-\varepsilon_{m})]\}
\label{eq:b_mk1}%
\end{align}
where $P$ denotes the principal value. Substituting the last result into the
corresponding terms containing $\hat{V}_{M}$ on the right-hand side of
Eqs.(\ref{eq:n_m}) and (\ref{eq:p_M4}) and keeping only resonant terms, we have%

\begin{align}
\frac{2}{\hbar}\operatorname{Im}\sum_{K=L,R}\sum_{k\in K}V_{km}^{(MK)}\langle
b_{mk}\rangle &  =\sum_{K=L,R}[n_{m}(t)\Gamma_{MK,m}-\nonumber\\
&  -W_{MK,m}] \label{eq:n_m_MKrelaxation}%
\end{align}
and%
\begin{align}
&  \frac{i}{\hbar}\sum_{K=L,R}\sum_{k\in K}(V_{k1}^{(MK)}\langle b_{k2}%
^{+}\rangle-V_{2k}^{(MK)}\langle b_{k1}\rangle)=\nonumber\\
&  =-p_{M}(t)\sum_{K=L,R}[\frac{1}{2}(\Gamma_{MK,1}+\Gamma_{MK,2}%
)+i\Delta_{MK}] \label{eq:p_MKrelaxation}%
\end{align}
where%

\begin{equation}
\Gamma_{MK,m}=\frac{2\pi}{\hbar}\sum_{k\in K}|V_{km}^{(MK)}|^{2}%
\delta(\varepsilon_{k}-\varepsilon_{m}), \label{eq:GammaMKm}%
\end{equation}%
\begin{equation}
\Delta_{MK}=\frac{1}{\hbar}P\sum_{k\in K}[\frac{|V_{k1}^{(MK)}|^{2}%
}{\varepsilon_{k}-\varepsilon_{1}}-\frac{|V_{2k}^{(MK)}|^{2}}{\varepsilon
_{k}-\varepsilon_{2}}] \label{eq:Im(MKterm_for_pM)}%
\end{equation}
is the correction to the frequency of molecular transition $(\varepsilon
_{2}-\varepsilon_{1})/\hbar$ due to electron transfer between the molecule and
lead $K$,
\begin{align}
W_{MK,m}  &  =\frac{2\pi}{\hbar}\sum_{k\in K}f_{K}(\varepsilon_{k}%
)|V_{km}^{(MK)}|^{2}\delta(\varepsilon_{k}-\varepsilon_{m})\label{eq:WMKm}\\
&  =f_{K}(\varepsilon_{m})\Gamma_{MK,m} \label{eq:WMKm1}%
\end{align}
It should be noted that the second equality, Eq.(\ref{eq:WMKm1}), is valid
only provided that molecular state $\varepsilon_{m}$ is far from the Fermi
level of lead $K$. The point is that the position of $\varepsilon_{m}$
influences on the bath correlation frequency, $\omega_{c}$, for the relaxation
parameter $W_{MK,m}$ ($\omega_{c}$ is the frequency interval at which the
interaction of a system (molecule) with bath (metal leads) is essentially
changed). Really, if molecular states $\varepsilon_{m}$ are far from the Fermi
levels of both leads, $\omega_{c}$ for $W_{MK,m}$ is the same as that for
$\Gamma_{MK,m}$, which is determined by the frequency interval for the
system-bath interaction matrix element $V_{km}^{(MK)}$ and the density of
states of metal leads. For this case $\omega_{c}$ can be evaluated as 1-10 eV
\cite{Schreiber06}. The situation is different if we assume that the molecular
level position is pinned to the Fermi energy of a lead. In the latter case
$\omega_{c}$ for $W_{MK,m}$ is determined also by the frequency interval at
which $f_{K}(\varepsilon)$ is essentially changed that is $\sim k_{B}%
T/\hbar=0.026$ eV for room temperature (see Eq.(\ref{eq:WMKm})). Since the
value of $\omega_{c}$ places a limit on the used approximation, according to
which the relaxation parameters do not depend on exciting electromagnetic
field (see Sec.\ref{sec:Conclusion}), one can use Eq.(\ref{eq:WMKm1}) only in
the case when the bath correlation frequency $\omega_{c}$ is the same for both
$W_{MK,m}$ and $\Gamma_{MK,m}$.

One can easily see from Eqs.(\ref{eq:n_m}), (\ref{eq:p_M4}),
(\ref{eq:n_m_MKrelaxation}) and (\ref{eq:p_MKrelaxation}) that in the absence
of energy transfer ($\hat{V}_{N}$), equations for the populations of molecular
states and molecular polarization form a closed set of the equations of motion.

\subsection{Calculation of terms related to energy transfer in the equations
for $n_{m}$ and $p_{M}$}

The calculation of terms related to energy transfer in Eqs.(\ref{eq:n_m}) and
(\ref{eq:p_M4}) is similar to that of
Sec.\ref{subsec:Mkrelaxation_for_n_and_p}. Invoking again the non-crossing
approximation by omitting $\hat{V}_{P}$ and $\hat{V}_{M}$ terms in the
equations of motion for the expectation values $\langle b_{M}b_{k^{\prime}%
k}^{+}\rangle$ and $\langle b_{k^{\prime}k}(\hat{n}_{2}-\hat{n}_{1})\rangle$,
which appear on the right-hand sides of Eqs.(\ref{eq:n_m}) and (\ref{eq:p_M4}%
), respectively, we get%
\begin{align}
\frac{d}{dt}\langle b_{M}b_{k^{\prime}k}^{+}\rangle &  =\frac{i}{\hbar
}(\varepsilon_{k^{\prime}}-\varepsilon_{k}-\varepsilon_{2}+\varepsilon
_{1})\langle b_{M}b_{k^{\prime}k}^{+}\rangle+\nonumber\\
&  +\frac{i}{\hbar}V_{kk^{\prime}}^{(NK)}\{f_{K}(\varepsilon_{k}%
)[1-f_{K}(\varepsilon_{k^{\prime}})]\langle b_{M}^{+}b_{M}\rangle-\nonumber\\
&  -f_{K}(\varepsilon_{k^{\prime}})[1-f_{K}(\varepsilon_{k})]\langle
b_{M}b_{M}^{+}\rangle\}
\end{align}%
\begin{align}
\frac{d}{dt}\langle b_{k^{\prime}k}(\hat{n}_{2}-\hat{n}_{1})\rangle &
=\frac{i}{\hbar}(\varepsilon_{k}-\varepsilon_{k^{\prime}})\langle
b_{k^{\prime}k}(\hat{n}_{2}-\hat{n}_{1})\rangle+\nonumber\\
&  +\frac{i}{\hbar}V_{k^{\prime}k}^{(NK)}\{f_{K}(\varepsilon_{k^{\prime}%
})[1-f_{K}(\varepsilon_{k})]+\nonumber\\
&  +f_{K}(\varepsilon_{k})[1-f_{K}(\varepsilon_{k^{\prime}})]\}p_{M}%
\end{align}
Formally integrating the last equations, performing Markovian approximation
and substituting the results into the corresponding terms containing $\hat
{V}_{N}$ on the right-hand side of Eqs.(\ref{eq:n_m}) and (\ref{eq:p_M4}), we obtain%

\begin{align}
&  -\frac{2}{\hbar}\operatorname{Im}\sum_{K=L,R}\sum_{k\neq k^{\prime}\in
K}V_{k^{\prime}k}^{(NK)}\langle b_{M}b_{k^{\prime}k}^{+}\rangle\nonumber\\
&  =\sum_{K=L,R}[B_{NK}(\varepsilon_{1}-\varepsilon_{2},\mu_{K})\langle
b_{M}b_{M}^{+}\rangle\nonumber\\
&  -B_{NK}(\varepsilon_{2}-\varepsilon_{1},\mu_{K})\langle b_{M}^{+}%
b_{M}\rangle] \label{eq:Im<b_Mb+_k'k>}%
\end{align}
and%
\begin{widetext}
\begin{equation}
\frac{i}{\hbar}\sum_{K=L,R}\sum_{k\neq k^{\prime}\in
K}V_{kk^{\prime} }^{(NK)}\langle
b_{k^{\prime}k}(\hat{n}_{2}-\hat{n}_{1})\rangle=-p_{M}(t)
\sum_{K=L,R}\{i\Delta_{NK}+\frac{1}{2}[B_{NK}(\varepsilon
_{1}-\varepsilon_{2},\mu_{K})+B_{NK}(\varepsilon_{2}-\varepsilon_{1},\mu
_{K})]\} \label{eq:Vkk'<b_k'k(n2-n1)>}
\end{equation}
\end{widetext}

where%

\begin{align}
B_{NK}(\varepsilon_{m}-\varepsilon_{n},\mu_{K})  &  =\frac{2\pi}{\hbar}%
\sum_{k\neq k^{\prime}\in K}|V_{kk^{\prime}}^{(NK)}|^{2}\times\nonumber\\
\times\delta(\varepsilon_{k}-\varepsilon_{k^{\prime}}+\varepsilon
_{m}-\varepsilon_{n})  &  f_{K}(\varepsilon_{k})[1-f_{K}(\varepsilon
_{k^{\prime}})] \label{eq:BNK}%
\end{align}
and%
\begin{align}
\Delta_{NK}  &  =\frac{1}{\hbar}P\sum_{k\neq k^{\prime}\in K}\{f_{K}%
(\varepsilon_{k^{\prime}})[1-f_{K}(\varepsilon_{k})]+\nonumber\\
&  +f_{K}(\varepsilon_{k})[1-f_{K}(\varepsilon_{k^{\prime}})]\}\frac
{|V_{kk^{\prime}}^{(NK)}|^{2}}{\varepsilon_{k}-\varepsilon_{k^{\prime}%
}+\varepsilon_{2}-\varepsilon_{1}} \label{eq:deltaNK}%
\end{align}
is the correction to the frequency of molecular transition $(\varepsilon
_{2}-\varepsilon_{1})/\hbar$ due to energy transfer between the molecule and
lead $K$. In deriving Eqs.(\ref{eq:Im<b_Mb+_k'k>}) and
(\ref{eq:Vkk'<b_k'k(n2-n1)>}) we have used the arguments, which are similar to
those used above in the derivation of Eqs.(\ref{eq:n_m_MKrelaxation}) and
(\ref{eq:p_MKrelaxation}).

One can see from Eqs.(\ref{eq:n_m}) and (\ref{eq:Im<b_Mb+_k'k>}) that in the
presence of energy transfer ($\hat{V}_{N}$), equations for the populations of
molecular states and molecular polarization do not form a closed set of the
equations of motion. They must be supplemented, at least, with equations for
the expectation values of tetradic variables $\langle b_{M}b_{M}^{+}\rangle$
and $\langle b_{M}^{+}b_{M}\rangle=N_{M}$, where $\langle b_{M}b_{M}%
^{+}\rangle$ and $N_{M}$ are related by the following equation:%
\begin{equation}
\langle b_{M}b_{M}^{+}\rangle=N_{M}-n_{2}+n_{1}
\label{eq:<b_Mb+_M>and<b+_Mb_M>}%
\end{equation}

\subsection{Equation for $\langle b_{M}b_{M}^{+}\rangle$}

Using Eq.(\ref{eq:Heis1}), straightforward operator algebra manipulations
yield for $\langle b_{M}b_{M}^{+}\rangle$ in RWA
\begin{align}
\frac{d\langle b_{M}b_{M}^{+}\rangle}{dt}  &  =-\frac{2}{\hbar}%
\operatorname{Im}\sum_{K=L,R}\sum_{k\in K}(V_{k1}^{(MK)}\langle b_{2k}%
b_{M}^{+}\rangle-\nonumber\\
&  -V_{2k}^{(MK)}\langle b_{k1}b_{M}^{+}\rangle)+\nonumber\\
&  +\frac{2}{\hbar}\operatorname{Im}\sum_{K=L,R}\sum_{k\neq k^{\prime}\in
K}V_{k^{\prime}k}^{(NK)}\langle b_{M}b_{k^{\prime}k}^{+}\rangle-\nonumber\\
&  -\operatorname{Im}\{\Omega^{\ast}(t)\exp[i\omega t-i\varphi(t)]p_{M}\}
\label{eq:d<b_Mb+_M>}%
\end{align}
where the second term on the right-hand side of Eq.(\ref{eq:d<b_Mb+_M>}) has
been calculated above, Eq.(\ref{eq:Im<b_Mb+_k'k>}). The first term on the
right-hand sides of Eq.(\ref{eq:d<b_Mb+_M>}) is associated with the electron
transfer process. To evaluate it we consider the equations of motion for the
expectation values $\langle b_{2k}b_{M}^{+}\rangle$ and $\langle b_{k1}%
b_{M}^{+}\rangle$, omitting $\hat{V}_{P}$ and $\hat{V}_{N}$ interactions and
keeping only resonant terms:
\begin{align}
\frac{d\langle b_{2k}b_{M}^{+}\rangle}{dt}  &  =\frac{i}{\hbar}(\varepsilon
_{k}-\varepsilon_{1})\langle b_{2k}b_{M}^{+}\rangle+\frac{i}{\hbar}%
V_{1k}^{(MK)}[\langle b_{M}b_{M}^{+}\rangle-\nonumber\\
&  -f_{K}(\varepsilon_{k})(1-n_{2})] \label{eq:<b_2k_b+_M>}%
\end{align}%
\begin{eqnarray}
\frac{d\langle b_{k1}b_{M}^{+}\rangle}{dt}
=\frac{i}{\hbar}(\varepsilon
_{2}-\varepsilon_{k})\langle b_{k1}b_{M}^{+}\rangle+\nonumber\\
+\frac{i}{\hbar}V_{k2}^{(MK)}[(1-f_{K}(\varepsilon_{k}))n_{1}
-\langle b_{M}b_{M}^{+}\rangle] \label{eq:<b_k1_b+_M>}%
\end{eqnarray}
Integrating Eqs.(\ref{eq:<b_2k_b+_M>}) and (\ref{eq:<b_k1_b+_M>}),
performing Markovian approximation and substituting the results into
the first term on
the right-hand side of Eq.(\ref{eq:d<b_Mb+_M>}), we get%

\begin{align}
&  -\frac{2}{\hbar}\operatorname{Im}\sum_{K=L,R}\sum_{k\in K}(V_{k1}%
^{(MK)}\langle b_{2k}b_{M}^{+}\rangle-V_{2k}^{(MK)}\langle b_{k1}b_{M}%
^{+}\rangle)=\nonumber\\
&  =\sum_{K=L,R}[-\langle b_{M}b_{M}^{+}\rangle(\Gamma_{MK,1}+\Gamma
_{MK,2})+(1-n_{2})W_{MK,1}+\nonumber\\
&  +(\Gamma_{MK,2}-W_{MK,2})n_{1}] \label{eq:<b_Mb+_M>Rel}%
\end{align}
where $\Gamma_{MK,m}$ and $W_{MK,m}$ were defined in Eqs.(\ref{eq:GammaMKm}),
(\ref{eq:WMKm}) and (\ref{eq:WMKm1}).

\subsection{A closed set of the equations of motion}

Using Eqs.(\ref{eq:n_m}), (\ref{eq:p_M4}), (\ref{eq:n_m_MKrelaxation}),
(\ref{eq:p_MKrelaxation}), (\ref{eq:GammaMKm}), (\ref{eq:Im(MKterm_for_pM)}),
(\ref{eq:WMKm}), (\ref{eq:WMKm1}), (\ref{eq:Im<b_Mb+_k'k>}),
(\ref{eq:Vkk'<b_k'k(n2-n1)>}), (\ref{eq:BNK}), (\ref{eq:deltaNK}),
(\ref{eq:<b_Mb+_M>and<b+_Mb_M>}), (\ref{eq:d<b_Mb+_M>}),
(\ref{eq:<b_Mb+_M>Rel}) and switching to the system that rotates with
instantaneous frequency $\omega(t)$, $\tilde{p}_{M}(t)=p_{M}(t)\exp\{i[\omega
t-\varphi(t)]\}$, we obtain a closed set of equations for the quantities that
vary slowly with time during the period of a light wave%
\begin{widetext}
\begin{equation}
\frac{dn_{m}}{dt}=(-1)^{m}\operatorname{Im}[\Omega^{\ast}(t)\tilde{p}
_{M}]+W_{Mm}-\Gamma_{Mm}n_{m}-(-1)^{m}\lbrack
B_{N}(\varepsilon_{2}-\varepsilon_{1})N_{M}-(n_{1}
-n_{2}+N_{M})B_{N}(\varepsilon_{1}-\varepsilon_{2})]
\label{eq:n_m1a}
\end{equation}
\end{widetext}
\begin{equation}
\frac{d\tilde{p}_{M}}{dt}=i[\omega(t)-\omega_{0}]\tilde{p}_{M}+\frac{i}%
{2}\Omega(t)(n_{1}-n_{2})-\frac{1}{2}\Gamma_{MN}\tilde{p}_{M} \label{eq:p_M6}%
\end{equation}%
\begin{align}
\frac{dN_{M}}{dt}  &  =(\Gamma_{M1}-W_{M1})n_{2}+W_{M2}(1-n_{1}%
)+\operatorname{Im}[\Omega^{\ast}(t)\tilde{p}_{M}]-\nonumber\\
&  -[B_{N}(\varepsilon_{2}-\varepsilon_{1})+\tilde{\Gamma}_{M1}+\tilde{\Gamma
}_{M2}]N_{M}+\nonumber\\
&  +(n_{1}-n_{2}+N_{M})B_{N}(\varepsilon_{1}-\varepsilon_{2}) \label{eq:N_M}%
\end{align}
where%

\begin{equation}
\Gamma_{Mm}=\sum_{K=L,R}\Gamma_{MK,m},\text{ }W_{Mm}=\sum_{K=L,R}W_{MK,m}
\label{eq:GammaMm}%
\end{equation}

\begin{equation}
\sum_{K=L,R}B_{NK}(\varepsilon_{m}-\varepsilon_{n},\mu_{K})=B_{N}%
(\varepsilon_{m}-\varepsilon_{n}) \label{eq:BN}%
\end{equation}%
\begin{equation}
\Gamma_{MN}=\Gamma_{M1}+\Gamma_{M2}+B_{N}(\varepsilon_{2}-\varepsilon
_{1})+B_{N}(\varepsilon_{1}-\varepsilon_{2}) \label{eq:GammaMN}%
\end{equation}
and $\omega_{0}\equiv(\varepsilon_{2}-\varepsilon_{1})/\hbar+\sum
_{K=L,R}(\Delta_{NK}+\Delta_{MK})$ is the frequency of the molecular
transition with the corrections due to energy and electron transfer between
the molecule and the leads.

As indicated above, equations for the populations of molecular states and
molecular polarization, Eqs.(\ref{eq:n_m1a}) and (\ref{eq:p_M6}), form a
closed set of the equations of motion if the energy transfer is absent
($B_{N}(\varepsilon_{m}-\varepsilon_{n})=0$). When energy transfer is present
they must be supplemented with Eq.(\ref{eq:N_M}) for the exciton population.
On the other hand, in the absence of electron transfer ($W_{Mm}=\Gamma_{Mm}%
=0$) Eq.(\ref{eq:N_M}) coincides with Eq.(\ref{eq:n_m1a}) for $n_{2}$, which
implies that $N_{M}=n_{2}$. Indeed, $N_{M}=\langle\hat{c}_{2}^{+}\hat{c}%
_{1}\hat{c}_{1}^{+}\hat{c}_{2}\rangle=\langle\hat{n}_{2}(1-\hat{n}_{1}%
)\rangle=\langle\hat{n}_{2}^{2}\rangle=\langle\hat{n}_{2}\rangle=n_{2}$ when
the electron population on the molecule is conserved, i.e. when $\hat{V}%
_{M}=0$. It is the combined effect of the electron and energy transfer,
represented by the terms $\hat{V}_{M}$ and $\hat{V}_{N}$ in the Hamiltonian,
that leads to the need to include Eq.(\ref{eq:N_M}) in the closed set of the
equations of motion.

\subsection{Calculation of current and transferred charge}

The electronic current $I$ is given by the rate at which the number of
electrons changes in any of leads, e.g. \cite{Schreiber06,Harbola_Mukamel06}
\begin{equation}
I=e\frac{d}{dt}\sum_{k\in L}\langle\hat{n}_{k}\rangle=\frac{ie}{\hbar}%
\sum_{k\in L}\langle\lbrack\hat{H},\hat{n}_{k}]\rangle\label{eq:I}%
\end{equation}
Evaluating the commutator in Eq.(\ref{eq:I}), we get%

\begin{align}
I  &  =\frac{2e}{\hbar}\operatorname{Im}\sum_{m=1,2}\sum_{k\in L}V_{km}%
^{(MK)}\langle b_{mk}\rangle\nonumber\\
&  =e\sum_{m=1,2}[n_{m}(t)\Gamma_{ML,m}-W_{ML,m}] \label{eq:I_1}%
\end{align}
where we used Eq.(\ref{eq:n_m_MKrelaxation}). Correspondingly, the charge
transferred during an electromagnetic pulse of finite duration is given by
$Q=\int_{-\infty}^{\infty}I(t)dt$.

In the Appendix A we show that in the absence of the radiative and
nonradiative energy transfer couplings, $\hat{V}_{P}$ and $\hat{V}_{N}$,
Eqs.(\ref{eq:GammaMKm}), (\ref{eq:WMKm}), (\ref{eq:WMKm1}), (\ref{eq:n_m1a}),
(\ref{eq:GammaMm}) and (\ref{eq:I_1}) lead to the well known Landauer formula
for the current \cite{Hau96}.

\section{Current induced by a quasistationary light pulse}

\label{sec:Strong_field}

In this section we calculate the current induced in a molecular nanojunctions
by a strong quasistationary light pulse. Here and in the next section we
assume that the molecular energy gap $\varepsilon_{2}-\varepsilon_{1}$ is much
larger than the voltage bias $\mu_{L}-\mu_{R}$ and that the HOMO and LUMO
energies, $\varepsilon_{1}$ and $\varepsilon_{2}$, are positioned rather far
($\gg k_{B}T$) from the Fermi levels of both leads, so that the dark
(Landauer) current through the junction is small and may be disregarded. Using
for this situation $W_{MK,1}\approx\Gamma_{MK,1},$ $W_{MK,2}\approx
B_{NK}(\varepsilon_{1}-\varepsilon_{2},\mu_{K})\approx0$, we obtain from
Eqs.(\ref{eq:n_m1a}), (\ref{eq:p_M6}), (\ref{eq:N_M}) and (\ref{eq:I_1})%

\begin{equation}
\frac{dn_{1}}{dt}=-\operatorname{Im}[\Omega^{\ast}(t)\tilde{p}_{M}%
]+\Gamma_{M1}(1-n_{1})+B_{N}(\varepsilon_{2}-\varepsilon_{1})N_{M}
\label{eq:n_1}%
\end{equation}%
\begin{equation}
\frac{dn_{2}}{dt}=\operatorname{Im}[\Omega^{\ast}(t)\tilde{p}_{M}]-\Gamma
_{M2}n_{2}-B_{N}(\varepsilon_{2}-\varepsilon_{1})N_{M} \label{eq:n_2}%
\end{equation}%
\begin{equation}
\frac{d\tilde{p}_{M}}{dt}=i[\omega(t)-\omega_{0}]\tilde{p}_{M}+\frac{i}%
{2}\Omega(t)(n_{1}-n_{2})-\frac{1}{2}\Gamma_{MN}\tilde{p}_{M} \label{eq:p_M}%
\end{equation}%
\begin{equation}
\frac{dN_{M}}{dt}=\operatorname{Im}[\Omega^{\ast}(t)\tilde{p}_{M}%
]-[B_{N}(\varepsilon_{2}-\varepsilon_{1})+\Gamma_{M1}+\Gamma_{M2}]N_{M}
\label{eq:N_M1}%
\end{equation}

\begin{equation}
I=e[(n_{1}-1)\Gamma_{ML,1}+n_{2}\Gamma_{ML,2}] \label{eq:I_2}%
\end{equation}
One can see from Eq.(\ref{eq:I_2}) that the current strongly increases when
$n_{2},1-n_{1}\sim1$, which can be realized for strong light fields. If we
further assume that the pulse amplitude $\mathcal{E}(t)$ and frequency
$\omega(t)$ change slowly on the time scale of all relaxation times as well as
the reciprocal Rabi frequency, one can put all time derivatives on the
left-hand sides of Eqs.(\ref{eq:n_1}), (\ref{eq:n_2}), (\ref{eq:p_M}) and
(\ref{eq:N_M1}) equal to zero, and the resulting stationary equations can be
easily solved
\begin{equation}
n_{2}=\frac{\Omega^{2}(t)(\Gamma_{M2}+\Gamma_{M1})/(4\Gamma_{M2})}%
{\frac{(\Gamma_{M2}+\Gamma_{M1})^{2}}{4\Gamma_{M1}\Gamma_{M2}}\Omega
^{2}(t)+(\Gamma_{MN}/2)^{2}+[\omega_{0}-\omega(t)]^{2}}
\label{eq:n2Ad_solution}%
\end{equation}

\begin{equation}
n_{1}=1-n_{2}\frac{\Gamma_{M2}}{\Gamma_{M1}} \label{eq:n1Ad_solution}%
\end{equation}

\begin{equation}
N_{M}=\frac{\Gamma_{M2}}{\Gamma_{M1}+\Gamma_{M2}}n_{2}
\label{eq:N_M_Ad_solution}%
\end{equation}

\begin{equation}
\tilde{p}_{M}=\frac{\Omega(t)}{2}\frac{i(\Gamma_{MN}/2)-[\omega(t)-\omega
_{0}]}{\Omega^{2}(t)\frac{(\Gamma_{M2}+\Gamma_{M1})^{2}}{4\Gamma_{M1}%
\Gamma_{M2}}+(\Gamma_{MN}/2)^{2}+[\omega_{0}-\omega(t)]^{2}}
\label{eq:p_M_Ad_solution}%
\end{equation}
This solution corresponds to the molecular level and exciton populations as
well as polarization adiabatically following the optical pulse. Substituting
Eqs.(\ref{eq:n2Ad_solution}) and (\ref{eq:n1Ad_solution}) into
Eq.(\ref{eq:I_2}), we get%
\begin{widetext}
\begin{equation}
I(t)=e\frac{\Gamma_{M2}+\Gamma_{M1}}{4\Gamma_{M1}\Gamma_{M2}}\frac{\Omega
^{2}(t)(\Gamma_{ML,2}\Gamma_{MR,1}-\Gamma_{ML,1}\Gamma_{MR,2})}{\Omega
^{2}(t)\frac{(\Gamma_{M2}+\Gamma_{M1})^{2}}{4\Gamma_{M1}\Gamma_{M2}}%
+(\frac{\Gamma_{MN}}{2})^{2}+[\omega_{0}-\omega(t)]^{2}} \label{eq:I_Ad1}%
\end{equation}
\end{widetext}
At steady-state ($\omega(t)=\omega,$ $\Omega(t)=\Omega$) and small
fields,
$\Omega^{2}\frac{(\Gamma_{M2}+\Gamma_{M1})^{2}}{4\Gamma_{M1}\Gamma_{M2}}%
\ll(\Gamma_{MN}/2)^{2},$ this becomes
\begin{widetext}
\begin{equation}
I=\frac{e\Omega^{2}}{4}\frac{\Gamma_{M2}+\Gamma_{M1}}{(\Gamma_{MN}%
/2)^{2}+(\omega_{0}-\omega)^{2}}\frac{\Gamma_{ML,2}\Gamma_{MR,1}-\Gamma
_{ML,1}\Gamma_{MR,2}}{\Gamma_{M1}\Gamma_{M2}} \label{eq:I_Ad_weak_field}%
\end{equation}
\end{widetext}
The last equation is similar to Eq.(50) of Ref.\cite{Nitzan06JCP}
with the only difference that the latter corresponds to the
substitution of the sum $\Gamma_{M2}+\Gamma_{M1}$ on the right-hand
side of
Eq.(\ref{eq:I_Ad_weak_field}) by $\Gamma_{MN}=\Gamma_{M2}+\Gamma_{M1}%
+B_{N}(\varepsilon_{2}-\varepsilon_{1})>\Gamma_{M2}+\Gamma_{M1}$. The
difference may arise from the fact that Eq.(50) of Ref.\cite{Nitzan06JCP} is
obtained in the much used approximation of strong dephasing \cite{Chow97}
$N_{M}=\langle\hat{n}_{2}(1-\hat{n}_{1})\rangle\approx n_{2}(1-n_{1})$. For
small fields the latter term is of order $\Omega^{4},$ since $n_{2}%
,1-n_{1}\sim\Omega^{2}$. As a matter of fact, when the exciting field is weak,
one can neglect the term $B_{N}(\varepsilon_{2}-\varepsilon_{1})n_{2}%
(1-n_{1})\sim\Omega^{4}$ with respect to $\Gamma_{M1}(1-n_{1})\sim\Omega^{2}$
and $\Gamma_{M2}n_{2}\sim\Omega^{2}$ on the right-hand sides of
Eqs.(\ref{eq:n_1}) and (\ref{eq:n_2}), respectively. In other words, the
approximation of strong dephasing $N_{M}\approx n_{2}(1-n_{1})$ disregards the
depletion of state $2$ due to energy transfer for small fields, and therefore
results in some overestimation of the current. In contrast, our present
approach takes the tetradic variable $N_{M}$ into account exactly (in the
framework of the Markovian approximation) and does describe the depletion of
state $2$ due to energy transfer.

For strong fields and near resonance excitation, $\Omega^{2}(t)\frac
{(\Gamma_{M2}+\Gamma_{M1})^{2}}{4\Gamma_{M1}\Gamma_{M2}}\gg(\Gamma_{MN}%
/2)^{2},[\omega_{0}-\omega(t)]^{2}$, Eqs.(\ref{eq:n2Ad_solution}),
(\ref{eq:n1Ad_solution}), (\ref{eq:N_M_Ad_solution}) and
(\ref{eq:p_M_Ad_solution}) describe the saturation effect
\begin{equation}
n_{1}=n_{2}=\frac{\Gamma_{M1}}{\Gamma_{M2}+\Gamma_{M1}},
\label{eq:n_1and2_saturation}%
\end{equation}

\begin{equation}
N_{M}=\frac{\Gamma_{M1}\Gamma_{M2}}{(\Gamma_{M2}+\Gamma_{M1})^{2}},
\label{eq:N_Msaturation}%
\end{equation}%
\begin{equation}
\tilde{p}_{M}=\frac{2\Gamma_{M1}\Gamma_{M2}}{(\Gamma_{M2}+\Gamma_{M1})^{2}%
}\frac{i(\Gamma_{MN}/2)-[\omega(t)-\omega_{0}]}{\Omega(t)},
\label{eq:p_Msaturation}%
\end{equation}
and Eq.(\ref{eq:I_1}) gives%
\begin{equation}
I=e\frac{\Gamma_{ML,2}\Gamma_{MR,1}-\Gamma_{ML,1}\Gamma_{MR,2}}{\Gamma
_{M2}+\Gamma_{M1}} \label{eq:I_saturation}%
\end{equation}
Eqs.(\ref{eq:I_Ad1}) and Eq.(\ref{eq:I_saturation}) show that the optically
induced current increases linearly with the pulse intensity $\sim\Omega^{2}$
for weak fields, and saturates at the maximal value given by
Eq.(\ref{eq:I_saturation}), for strong fields. As is easy to see from
Eqs.(\ref{eq:n_1and2_saturation}) and (\ref{eq:N_Msaturation}), in the latter
case $N_{M}=n_{2}(1-n_{1})$, since the strong dephasing limit where $\tilde
{p}_{M}$ can be disregarded is realized under saturation effect (see
Eq.(\ref{eq:p_Msaturation})).

\section{Optical control of current and transferred charge with chirped
pulses}

\label{sec:Optical control}

In the previous section we have generalized the results of
Refs.\cite{Nit05,Nitzan06JCP} to the quasistationary strong electromagnetic
field limit. As mentioned in the introduction, future generations of optical
communication systems will employ coherent optical manipulations whose speed
greatly exceeds that of currently available electronic devices. We next
consider such coherent control processes.

Two well-known procedures based on a coherent excitation can, in principle,
produce complete population inversion in an ensemble of two-level atoms. One
of them is the $\pi$-pulse excitation \cite{All75}, which makes use of the
Rabi population oscillations. This approach has been successfully demonstrated
in atoms as well as semiconductor quantum dots, often referred to as
artificial atoms
\cite{Stievater01,Kamada_Takagahara01,Htoon_Takagahara02,Zrenner02Nature}. The
main disadvantage of the $\pi$-pulse excitation method is the requirement for
resonant light source and the need of precise control of the pulse area
\cite{Sho92}.

The second procedure, known as adiabatic rapid passage (ARP)
\cite{Mel94,Sho92,All75,Tre68,Vit01,Fai02JCP_2,Fai04JCP,Fai05JOSAB}, enables
us to transfer the entire population from ground $|1>$ to the excited $|2>$
electronic state. It is based on sweeping the pulse frequency through a
resonance. The mechanism of ARP can be explained by avoided crossing of
dressed (adiabatic) states%

\begin{align}
\Phi_{+}(t)  &  =\sin\vartheta(t)|1>+\cos\vartheta
(t)|2>\label{eq:dressed_states}\\
\Phi_{-}(t)  &  =\cos\vartheta(t)|1>-\sin\vartheta(t)|2>\nonumber
\end{align}
as a function of the instantaneous laser pulse frequency $\omega(t)$
\cite{Sho92}. Here the mixing angle $\vartheta(t)$ is defined (modulo $\pi$)
as $\vartheta=(1/2)\arctan\frac{\Omega(t)}{\omega_{0}-\omega(t)}$, where
$\Omega(t)$ is the Rabi frequency. During the excitation, the mixing angle
rotates clockwise from $\vartheta(-\infty)=\pi/2$ to $\vartheta(+\infty)=0$
and the composition \ of adiabatic states changes accordingly. In particular,
starting from state $|1>$, the system follows the adiabatic (dressed) state
$\Phi_{+}(t)$ and eventually ends up in state $|2>$ \cite{Vit01}. A scheme
based on ARP is robust since it is insensitive to pulse area and the precise
location of the resonance. Therefore, we shall focus in what follows on ARP as
a way to control optically induced charge transfer in molecular nanojunctions.
The application of our formalism to the coherent optoelectronic properties of
nanojunctions with quantum dots, using $\pi$-pulse excitation,
\cite{Zrenner02Nature}, will be analyzed elsewhere.

As a particular example we shall consider a light-induced charge transfer in
molecular nanojunctions, using linear chirped pulses $\omega(t)=\omega
-\bar{\mu}(t-t_{0})$ where $\bar{\mu}=d^{2}\varphi(t)/dt^{2}=const$.

\subsection{Numerical results}

\begin{figure}
[ptb]
\begin{center}
\includegraphics[width=3.4in]%
{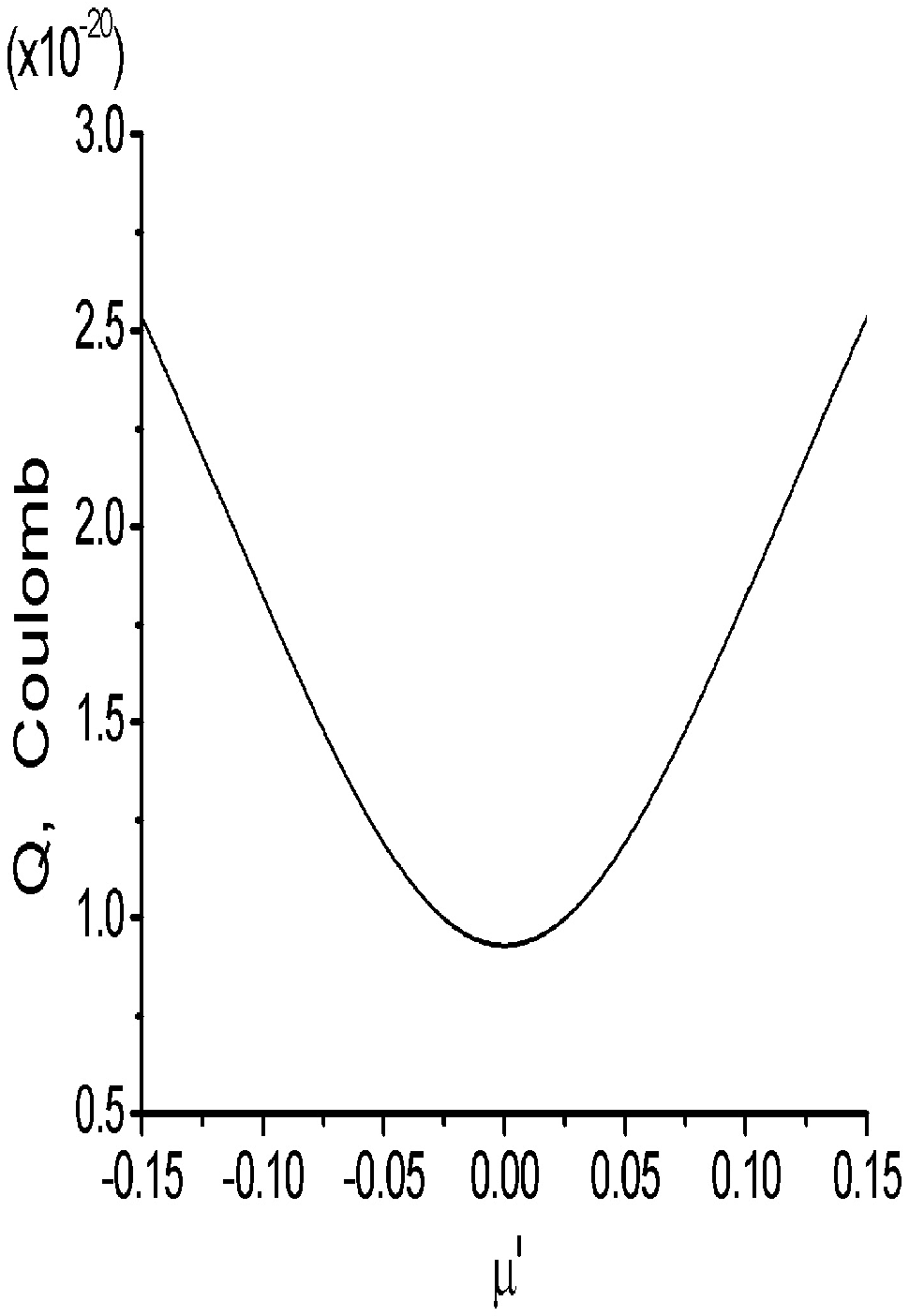}%
\caption{The charge $Q$ transferred after the completion of the
pulse action as a function of the linear chirp rate
$\mu^{\prime}=\bar{\mu}/\omega_{0}^{2}$ ($\omega_{0}$ is defined
below Eq.(\ref{eq:GammaMN})). Other parameters are as follows
$\omega=\omega_{0}=3$ eV, $\Gamma_{M2}/\omega_{0}=0.03$, $\Gamma
_{M1}/\omega_{0}=0.04$, $\Gamma_{MN}=\Gamma_{M1}+\Gamma_{M2}$, $d\mathcal{E}%
_{0}/\hbar\omega_{0}=0.2$, $\delta/\omega_{0}=0.1$, $\Gamma_{ML,1}=0.01$ eV,
$\Gamma_{ML,2}=0.02$ eV. }%
\label{fig:chirp_mu_a}%
\end{center}
\end{figure}

\begin{figure}
[ptb]
\begin{center}
\includegraphics[width=3.4in]%
{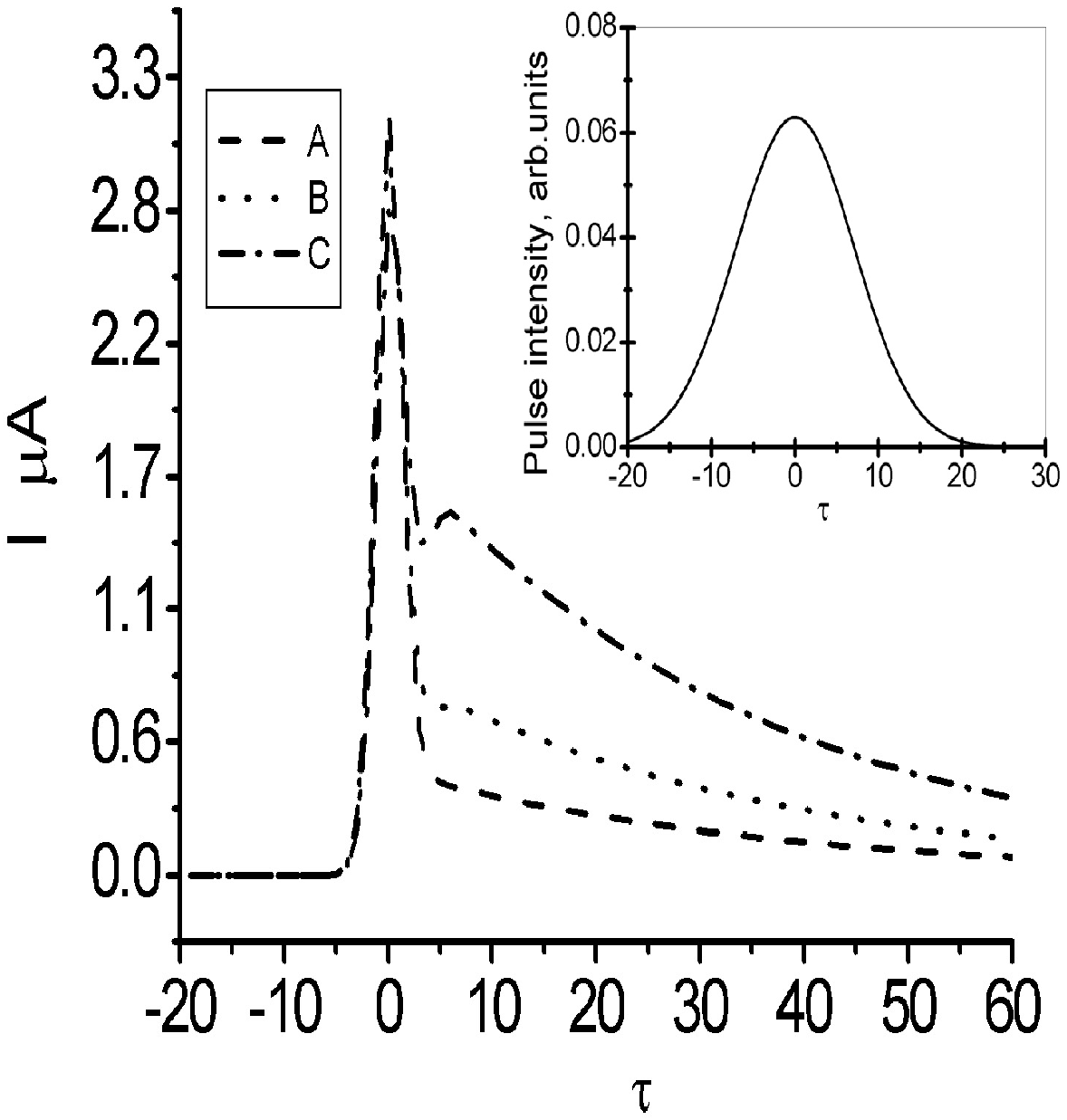}%
\caption{The current $I$ as a function of time $\tau=\omega_{0}t$
for the linear chirp rate $\bar{\mu}/\omega_{0}^{2}=0$ (A)$,$ $0.07$
(B) and $0.15$ (C). Other parameters are identical to those of
Fig.\ref{fig:chirp_mu_a}. The figure illustrates how signatures of
ARP increase the induced current. Inset, the square of electric
field amplitude of the exciting pulse in arbitrary
units. }%
\label{fig:chirp_curr_b}%
\end{center}
\end{figure}

Figures \ref{fig:chirp_mu_a} and \ref{fig:chirp_curr_b} show the influence of
$\bar{\mu}$, the chirp rate in the time domain, on the charge transferred
during one electromagnetic pulse action. These results are obtained by
numerical solution of Eqs.(\ref{eq:n_1}), (\ref{eq:n_2}), (\ref{eq:p_M}),
(\ref{eq:N_M1}) and (\ref{eq:I_2}) for a Gaussian pulse of the shape
\begin{equation}
E(t)\equiv\mathcal{E}\left(  t\right)  \exp\left(  i\varphi\left(  t\right)
\right)  =\mathcal{E}_{0}\exp[-\frac{1}{2}(\delta^{2}-i\bar{\mu})(t-t_{0}%
)^{2}] \label{eq:gausspulse}%
\end{equation}
and are displayed as a function of $\bar{\mu}$. We see that pulse chirping can
increase the transferred charge (Fig.\ref{fig:chirp_mu_a}) and the induced
current (Fig.\ref{fig:chirp_curr_b}) that can be explained by signatures of
ARP (see below).%

%TCIMACRO{\FRAME{ftbpFU}{2.8082in}{3.6535in}{0pt}{\Qcb{The charge $Q$
%transferred after the completion of the pulse action as a function of the
%linear chirp rate $\mu^{\prime}=\bar{\mu}/\omega_{0}^{2}$ ($\omega_{0}$ is
%defined below Eq.(\ref{eq:GammaMN})). Other parameters are as follows
%$\omega=\omega_{0}=3$ eV, $\Gamma_{M2}/\omega_{0}=0.03$, $\Gamma_{M1}%
%/\omega_{0}=0.04$, $\Gamma_{MN}=\Gamma_{M1}+\Gamma_{M2}$, $d\mathcal{E}%
%_{0}/\hbar\omega_{0}=0.2$, $\delta/\omega_{0}=0.1$, $\Gamma_{ML,1}=0.01$ eV,
%$\Gamma_{ML,2}=0.02$ eV. }}{\Qlb{fig:chirp_mu_a}}{fig_chirp_mu_a.eps}%
%width 2.8082in;  height 3.6535in;  depth 0pt;  original-width 4.7968in;
%original-height 6.2559in;  cropleft "0";  croptop "1";  cropright "1";
%cropbottom "0";  filename 'Fig_chirp_mu_a.eps';file-properties "XNPEU";}} }%
%BeginExpansion
%EndExpansion

%

%TCIMACRO{\FRAME{ftbpFU}{3.2614in}{3.39in}{0pt}{\Qcb{The current $I$ as a
%function of time $\tau=\omega_{0}t$ for the linear chirp rate $\bar{\mu
%}/\omega_{0}^{2}=0$ (A)$,$ $0.07$ (B) and $0.15$ (C). Other parameters are
%identical to those of Fig.\ref{fig:chirp_mu_a}. The figure illustrates how
%signatures of ARP increase the induced current. Inset, the square of electric
%field amplitude of the exciting pulse in arbitrary units. }}%
%{\Qlb{fig:chirp_curr_b}}{fig_chirp_curr_b.eps}%
%width 3.2614in;  height 3.39in;  depth 0pt;  original-width 5.3831in;
%original-height 5.5986in;  cropleft "0";  croptop "1";  cropright "1";
%cropbottom "0";  filename 'Fig_chirp_curr_b.eps';file-properties "XNPEU";}} }%
%BeginExpansion
%EndExpansion

If chirped pulses are obtained by changing the separation of pulse
compression gratings, the parameters $\delta$ and $\bar{\mu}$ are
determined by the formulae \cite{Cer96,Fai98}:
\begin{eqnarray}
\delta^{2}=2\tau_{p0}^{2}[\tau_{p0}^{4}+4\Phi^{\prime\prime2}\left(
\omega\right)  ]^{-1},\nonumber\\
\text{}\bar{\mu}=-4\Phi^{\prime\prime}\left( \omega\right)  \left[
\tau_{p0}^{4}+4\Phi^{\prime\prime2}\left( \omega\right)  \right]
^{-1}, \label{eq:deltamu}
\end{eqnarray}
where $\tau_{p0}=t_{p0}/\sqrt{2\ln2}$, $t_{p0}$ is the pulse
duration of the corresponding transform-limited pulse, and
$\Phi^{\prime\prime}\left( \omega\right)  $ is the chirp rate in the
frequency domain. The latter is defined by writing the electric
field at frequency $\tilde{\omega}$ as
$|E(\tilde{\omega})|\exp[i\Phi(\tilde{\omega})]$, and expanding
phase term $\Phi(\tilde{\omega})$ in a Taylor series about the
carrier frequency $\omega$
$\Phi(\tilde{\omega})=\Phi(\omega)+(1/2)\Phi^{\prime\prime}(\omega
)(\tilde{\omega}-\omega)^{2}+...$ Note that the local field in the
junction reflects also plasmon excitation in the leads, and taking
the incident pulse shape as affected only by the compression
gratings used disregards the possible contribution of the near field
responce of plasmonic excitations in
the leads \cite{Wang_Shen06,Brixner06}. Such effects will be considered elsewhere.%

%TCIMACRO{\FRAME{ftbpFU}{4.2602in}{3.3475in}{0pt}{\Qcb{The charge transferred
%after the completion of the pulse action as a function of the chirp rate in
%the frequency domain $\Phi^{\prime\prime}\left(  \nu\right)  $. The value of
%$d\mathcal{E}_{0}/\hbar\omega_{0}=0.6$ (A), $0.5$ (B), $0.4$ (C) and $0.3$ (D)
%for the transform-limited pulse. In the course of chirping the pulse energy is
%conserved so that $\int_{-\infty}^{\infty}\mathcal{E}^{2}(t)dt=\mathcal{E}%
%_{0}^{2}\sqrt{\frac{\pi}{2}[\tau_{p0}^{2}+\frac{4\Phi^{\prime\prime2}\left(
%\omega\right)  }{\tau_{p0}^{2}}]}$ = $const$, and $\mathcal{E}_{0}$ decreases
%when $\left\vert \Phi^{\prime\prime}\left(  \nu\right)  \right\vert $
%increases; $\tau_{p0}=11$ fs. Other parameters are identical to those of
%Fig.\ref{fig:chirp_mu_a}.}}{\Qlb{fig:chirp_Fi}}{fig_chirp_fi.eps}%
%{\special{ language "Scientific Word";  type "GRAPHIC";  display "USEDEF";
%valid_file "F";  width 4.2602in;  height 3.3475in;  depth 0pt;
%original-width 4.4269in;  original-height 3.6677in;  cropleft "0";
%filename 'Fig_chirp_Fi.eps';file-properties "XNPEU";}} }%
%BeginExpansion
\begin{figure}
[ptb]
\begin{center}
\includegraphics[width=3.4in]%
{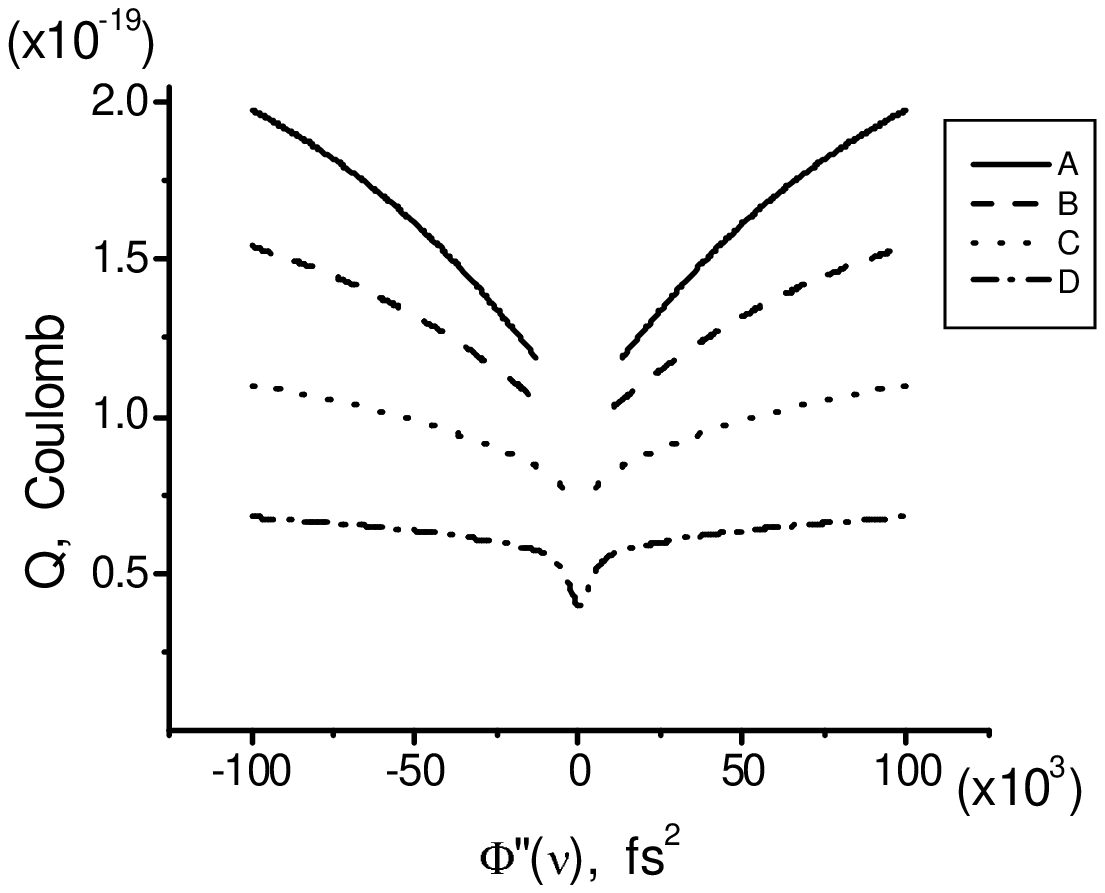}%
\caption{The charge transferred after the completion of the pulse
action as a function of the chirp rate in the frequency domain
$\Phi^{\prime\prime}\left( \nu\right)  $. The value of
$d\mathcal{E}_{0}/\hbar\omega_{0}=0.6$ (A), $0.5$ (B), $0.4$ (C) and
$0.3$ (D) for the transform-limited pulse. In the course of chirping
the pulse energy is conserved so that $\int_{-\infty}^{\infty
}\mathcal{E}^{2}(t)dt=\mathcal{E}_{0}^{2}\sqrt{\frac{\pi}{2}[\tau_{p0}%
^{2}+\frac{4\Phi^{\prime\prime2}\left(  \omega\right)  }{\tau_{p0}^{2}}]}$ =
$const$, and $\mathcal{E}_{0}$ decreases when $\left\vert \Phi^{\prime\prime
}\left(  \nu\right)  \right\vert $ increases; $\tau_{p0}=11$ fs. Other
parameters are identical to those of Fig.\ref{fig:chirp_mu_a}.}%
\label{fig:chirp_Fi}%
\end{center}
\end{figure}
%EndExpansion
%

%TCIMACRO{\FRAME{ftbpFU}{3.8114in}{3.0051in}{0pt}{\Qcb{The charge transferred
%after the completion of the pulse action as a function of the chirp rate in
%the frequency domain $\Phi^{\prime\prime}\left(  \nu\right)  $ in the presence
%of energy transfer $B_{N}(\varepsilon_{2}-\varepsilon_{1})/\omega_{0}=0.01$.
%Other parameters are identical to those of Fig.\ref{fig:chirp_Fi}. }%
%}{\Qlb{fig:chirp_Fi_4corr}}{figchirp_fi_4corr.eps}%
%width 3.8114in;  height 3.0051in;  depth 0pt;  original-width 5.0434in;
%original-height 3.9692in;  cropleft "0";  croptop "1";  cropright "1";
%cropbottom "0";  filename 'FigChirp_Fi_4Corr.eps';file-properties "XNPEU";}}
%}%
%BeginExpansion
\begin{figure}
[ptb]
\begin{center}
\includegraphics[width=3.4in]%
{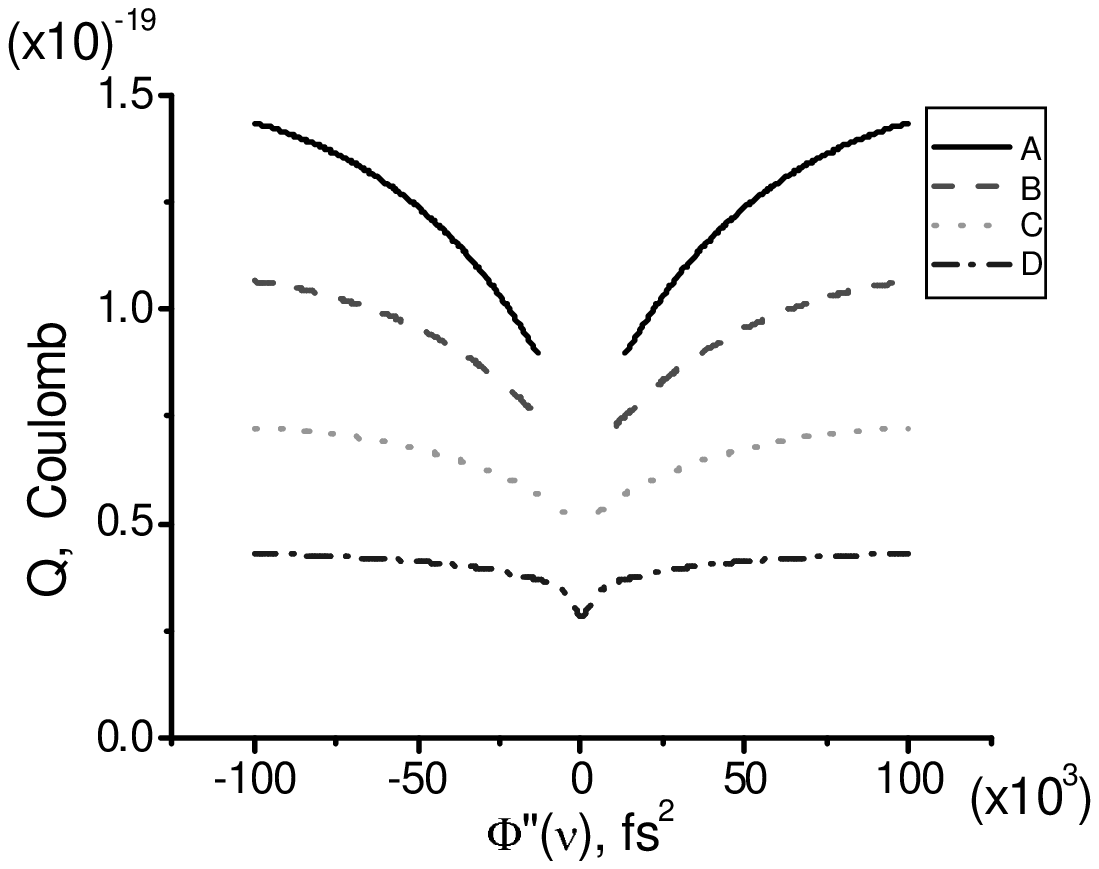}%
\caption{The charge transferred after the completion of the pulse
action as a function of the chirp rate in the frequency domain
$\Phi^{\prime\prime}\left( \nu\right)  $ in the presence of energy
transfer $B_{N}(\varepsilon _{2}-\varepsilon_{1})/\omega_{0}=0.01$.
Other parameters are identical to
those of Fig.\ref{fig:chirp_Fi}. }%
\label{fig:chirp_Fi_4corr}%
\end{center}
\end{figure}
\begin{figure}
[ptbptb]
\begin{center}
\includegraphics[width=3.4in]%
{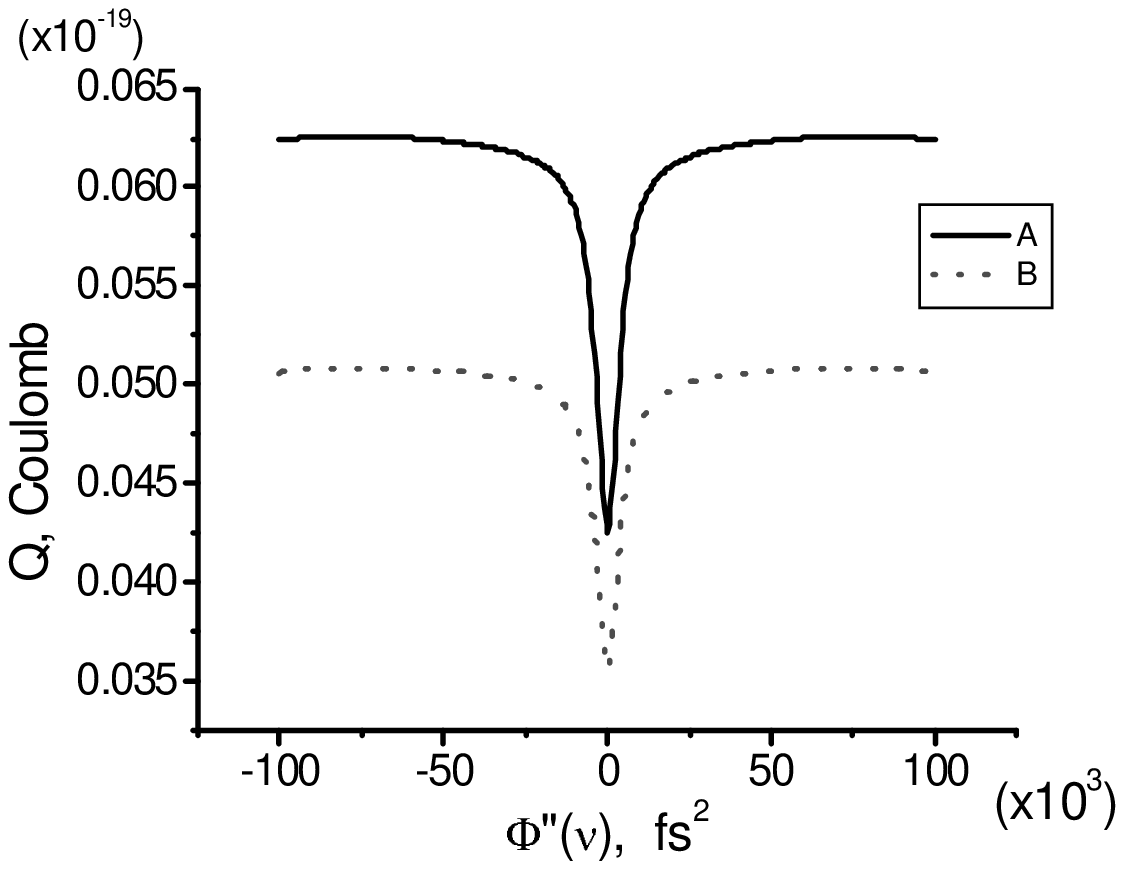}%
\caption{The charge transferred after the completion of the pulse
action as a function of the chirp rate in the frequency domain
$\Phi^{\prime\prime}\left( \nu\right)  $ when
$d\mathcal{E}_{0}/\hbar\omega_{0}=0.1$ in the absence
($B_{N}(\varepsilon_{2}-\varepsilon_{1})/\omega_{0}=0$, curve A) and
presence ($B_{N}(\varepsilon_{2}-\varepsilon_{1})/\omega_{0}=0.01$,
curve B) of energy
transfer. Other parameters are identical to those of Fig.\ref{fig:chirp_Fi}. }%
\label{fig:chirp_Fi_comparison}%
\end{center}
\end{figure}

%EndExpansion
Figs.\ref{fig:chirp_Fi} and \ref{fig:chirp_Fi_4corr} show the calculation
results of the transferred charge $Q$ as a function of the chirp rate in the
frequency domain $\Phi^{\prime\prime}\left(  \nu\right)  =4\pi^{2}\Phi
^{\prime\prime}\left(  \omega\right)  $. The calculated dependences
$Q(\Phi^{\prime\prime}\left(  \nu\right)  )$ for curves A, B and C are
confined to the values of an argument $\left\vert \Phi^{\prime\prime}\left(
\nu\right)  \right\vert >0$ corresponding to $d\mathcal{E}_{0}/\hbar\omega
_{0}\leq0.3$ ($d$ is the molecular dipole moment, cf Eq.(\ref{eq:V_P_res})),
since our theory uses RWA. One can see that $Q$ grows rapidly for small
$\left\vert \Phi^{\prime\prime}\left(  \nu\right)  \right\vert .$ The growth
of $Q$ slows down for moderate $\left\vert \Phi^{\prime\prime}\left(
\nu\right)  \right\vert ,$ and then $Q$ tends to a constant value for large
$\left\vert \Phi^{\prime\prime}\left(  \nu\right)  \right\vert $. The larger
is pulse energy, the larger is the value of $\left\vert \Phi^{\prime\prime
}\left(  \nu\right)  \right\vert $, at which the growth of $Q$ slows down.
Fig.\ref{fig:chirp_Fi} corresponds to the absence of the energy transfer
($B_{N}(\varepsilon_{2}-\varepsilon_{1})=0$), and Fig.
\ref{fig:chirp_Fi_4corr} illustrates the influence of the energy transfer
($B_{N}(\varepsilon_{2}-\varepsilon_{1})\neq0$), which diminishes the
corresponding values of $Q$ (see also Fig.\ref{fig:chirp_Fi_comparison}).%
%TCIMACRO{\FRAME{ftbpFU}{3.8947in}{3.1177in}{0pt}{\Qcb{The charge transferred
%after the completion of the pulse action as a function of the chirp rate in
%the frequency domain $\Phi^{\prime\prime}\left(  \nu\right)  $ when
%$d\mathcal{E}_{0}/\hbar\omega_{0}=0.1$ in the absence ($B_{N}(\varepsilon
%_{2}-\varepsilon_{1})/\omega_{0}=0$, curve A) and presence ($B_{N}%
%(\varepsilon_{2}-\varepsilon_{1})/\omega_{0}=0.01$, curve B) of energy
%transfer. Other parameters are identical to those of Fig.\ref{fig:chirp_Fi}.
%}}{\Qlb{fig:chirp_Fi_comparison}}{figchirp_fi_comparison.eps}%
%width 3.8947in;  height 3.1177in;  depth 0pt;  original-width 5.1285in;
%original-height 4.097in;  cropleft "0";  croptop "1";  cropright "1";
%filename 'FigChirp_Fi_comparison.eps';file-properties "XNPEU";}} }%
%BeginExpansion
%EndExpansion
The behavior and values of $Q$ shown in Figs.\ref{fig:chirp_mu_a},
\ref{fig:chirp_Fi}, \ref{fig:chirp_Fi_4corr} and \ref{fig:chirp_Fi_comparison}
can be rationalized by the theoretical consideration below.

\begin{figure}
[ptb]
\begin{center}
\includegraphics[width=3.4in]%
{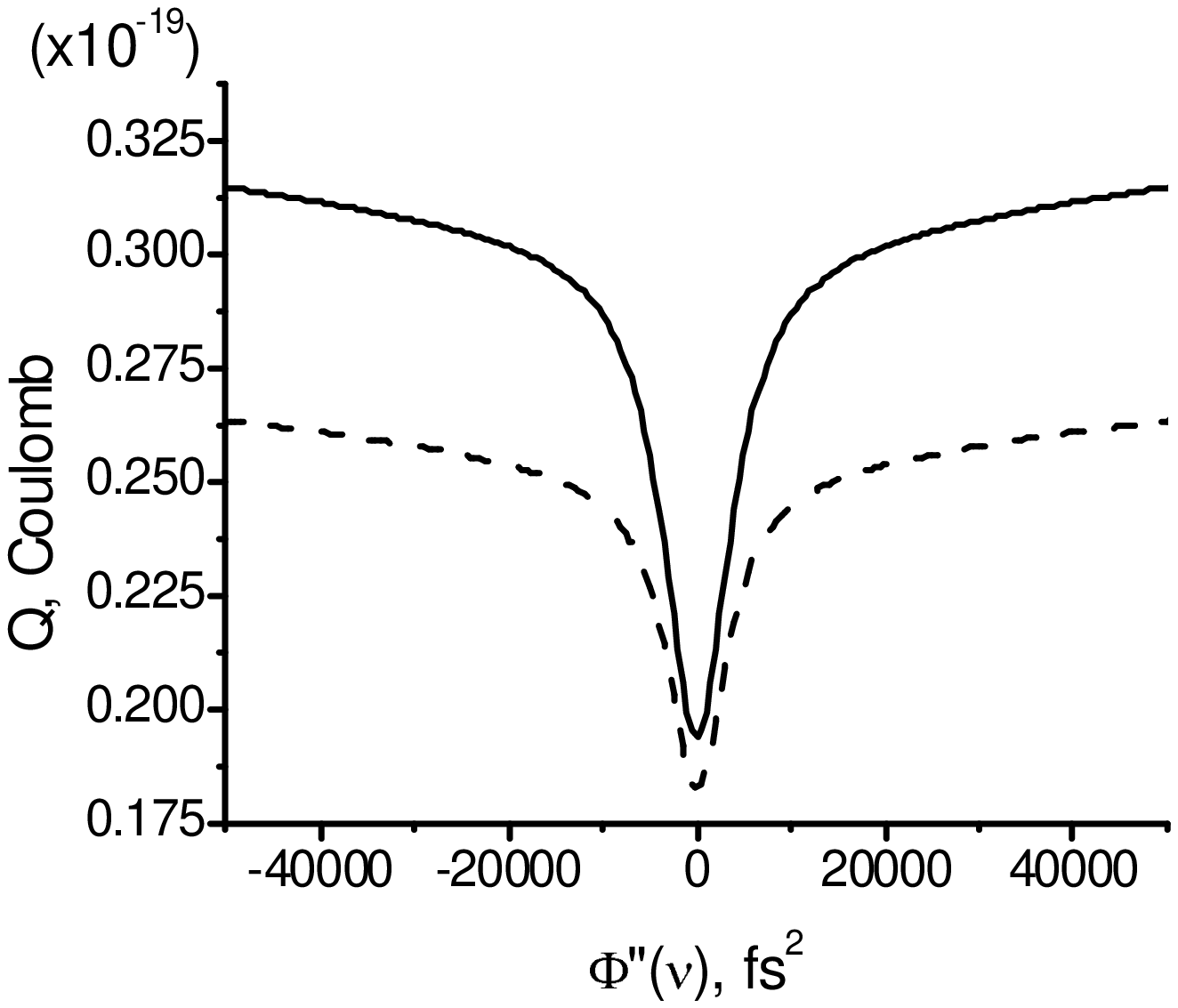}%
\caption{Influence of the frequency detuning $\omega_{0}-\omega$ on
the charge
transferred after the completion of the pulse action for $d\mathcal{E}%
_{0}/\hbar\omega_{0}=0.1$. $(\omega_{0}-\omega)/\omega_{0}=0$ (solid line) and
$0.05$ (dashed line). Other parameters are identical to those of
Fig.\ref{fig:chirp_Fi}. }%
\label{fig:chirp_Fi_detuning}%
\end{center}
\end{figure}

Fig. \ref{fig:chirp_Fi_detuning} illustrates the influence of detuning between
the carrier pulse frequency $\omega$ and the corrected frequency of the
molecular transition $\omega_{0}$ on the transferred charge $Q$.

%

%TCIMACRO{\FRAME{ftbpFU}{3.7732in}{3.1177in}{0pt}{\Qcb{Influence of the
%frequency detuning $\omega_{0}-\omega$ on the charge transferred after the
%completion of the pulse action for $d\mathcal{E}_{0}/\hbar\omega_{0}=0.1$.
%$(\omega_{0}-\omega)/\omega_{0}=0$ (solid line) and $0.05$ (dashed line).
%Other parameters are identical to those of Fig.\ref{fig:chirp_Fi}. }%
%}{\Qlb{fig:chirp_Fi_detuning}}{figchirp_fi_detuning.eps}%
%width 3.7732in;  height 3.1177in;  depth 0pt;  original-width 4.9999in;
%original-height 4.1254in;  cropleft "0";  croptop "1";  cropright "1";
%cropbottom "0";  filename 'FigChirp_Fi_detuning.eps';file-properties "XNPEU";}%
%} }%
%BeginExpansion
%EndExpansion

To end this section we note that the current that corresponds to the
expectation value of $Q=0.5\cdot10^{-19}$ $C$ per pulse (corresponding to
curve D in Fig.\ref{fig:chirp_Fi}) and to an estimated pulse repetition
frequency of 82 MHz \cite{Zrenner02Nature} results in a small but measurable
value of about $4\times10^{-12}$ ampere.

\subsection{Analytical consideration}

The problem under consideration above can be solved analytically in certain conditions.

\subsubsection{Chirped pulse control of charge transfer in molecular
nanojunctions as the Landau-Zener transition to a decaying level}

Consider first an excitation of the molecular nanojunction with a linear
chirped pulse $\omega(t)=\omega_{0}-\bar{\mu}t$ ($t_{0}=0,$ $\omega=\omega
_{0}$) of a constant amplitude ($\left\vert \Omega\right\vert =const$) in the
absence of energy transfer: $B_{N}(\varepsilon_{2}-\varepsilon_{1}%
)=\Delta_{NK}=0$, $\Gamma_{MN}=\Gamma_{M2}+\Gamma_{M1}$ ($\Gamma_{MN}$ was
defined by Eq.(\ref{eq:GammaMN})). If in addition, $\Gamma_{M1}=\Gamma
_{M2}\equiv\Gamma_{M}$ and provided that level $1$ is below and level $2$ is
above both Fermi energies, then it can be shown that $n_{1}=1-n_{2}$ (see
Appendix B) and
\begin{equation}
Q=e(\Gamma_{ML,2}-\Gamma_{ML,1})\int_{-\infty}^{\infty}n_{2}(t)dt
\label{eq:Q1}%
\end{equation}
Under these conditions our electron problem (Eqs.(\ref{eq:n_1})-(\ref{eq:N_M1}%
)) becomes mathematically equivalent to the Landau-Zener transition to a
decaying level\footnote{When the following corrections of misprints in
Ref.\cite{Aku92} are made: $Q=\rho_{12}+\rho_{21}$ and $\mathcal{P}%
=i(\rho_{21}-\rho_{12})$.} solved analytically by Akulin and Schleich
\cite{Aku92}. The magnitude $\Gamma_{ML,2}\int_{-\infty}^{\infty}n_{2}(t)dt$
on the right-hand side of Eq.(\ref{eq:Q1}) represents the expectation value of
the number of electrons passed from the molecule to the left lead after the
completion of the pulse action, and $\Gamma_{ML,1}\int_{-\infty}^{\infty}%
n_{2}(t)dt=\Gamma_{ML,1}\int_{-\infty}^{\infty}(1-n_{1}(t))dt$ is the same for
the electrons passed from the left lead to the molecule.

Using Eq.(\ref{eq:Q1}) and Eq.(25) of Ref.\cite{Aku92} for the magnitude
$I_{AS}\equiv$ $\Gamma_{M}\int_{-\infty}^{\infty}n_{2}(t)dt$, we obtain in
terms of our representation%

\begin{align}
Q  &  =e\frac{\Gamma_{ML,2}-\Gamma_{ML,1}}{\Gamma_{M}}I_{AS}=2\pi
e\frac{\Gamma_{ML,2}-\Gamma_{ML,1}}{\Gamma_{M}}\frac{\Omega^{2}}{4\left\vert
\bar{\mu}\right\vert }\times\nonumber\\
&  \times\exp\left(  \frac{-\pi\Omega^{2}}{4\left\vert \bar{\mu}\right\vert
}\right)  \left\vert W_{\frac{i\Omega^{2}}{4\left\vert \bar{\mu}\right\vert
},-1/2}(-\frac{i\Gamma_{M}^{2}}{\left\vert \bar{\mu}\right\vert })\right\vert
^{2} \label{eq:Qfinal}%
\end{align}
where $W_{ia,-1/2}(z)$ is the Whittaker function \cite{Magnus54}. The graph of
$I_{AS}$ as a function of Landau-Zener parameter and quenching parameter,
which correspond to $\Omega^{2}/\left\vert \bar{\mu}\right\vert $ and
$\Gamma_{M}^{2}/\left\vert \bar{\mu}\right\vert $, respectively, in terms of
our representation, can be found in Fig.1 of Ref.\cite{Aku92}.

When chirp is fast with respect to the rate of the electron transfer,
$\frac{\Gamma_{M}^{2}}{\left\vert \bar{\mu}\right\vert }\ll1$, one gets from
Eq.(\ref{eq:Qfinal})%

\begin{equation}
Q=e\frac{\Gamma_{ML,2}-\Gamma_{ML,1}}{\Gamma_{M}}\left[  1-\exp\left(
\frac{-\pi\Omega^{2}}{2\left\vert \bar{\mu}\right\vert }\right)  \right]  ,
\label{eq:Q_LZ}%
\end{equation}
where we have used the integral representation \cite{Magnus54} of the
Whittaker function to calculate $\left\vert \lim_{z\rightarrow0}%
W_{ia,-1/2}(z)\right\vert ^{2}=(a\pi)^{-1}\sinh(\pi a).$ The expression in the
brackets on the right-hand side of Eq.(\ref{eq:Q_LZ}) is simply the
probability of Landau-Zener transition, which indeed is identical to the
probability of the electron transfer from the excited molecule to the leads in
the case of fast passage through the resonance. Indeed, in this case
$\Gamma_{M}\int_{-\infty}^{\infty}n_{2}(t)dt=\Gamma_{M}\int_{0}^{\infty}%
n_{2}(0)\exp(-\Gamma_{M}t)dt=n_{2}(0)$ where $n_{2}(0)$ is the population of
molecular state $2$ immediately following the passage through the resonance.
The highest charge transfer is therefore obtained if $n_{2}(0)=1$.
Eq.(\ref{eq:Q_LZ}) shows that $n_{2}(0)$ approaches $1$ for strong
interaction, $\pi\Omega^{2}\gg2\left\vert \bar{\mu}\right\vert $, which
corresponds to adiabatic rapid passage (ARP). In other words, when the
interaction with light is short in comparison with the electron transfer, the
transferred charge is maximal when ARP is realized. Really, $Q\rightarrow|e|$
if $\left\vert \frac{\Gamma_{ML,2}-\Gamma_{ML,1}}{\Gamma_{M}}\right\vert
\rightarrow1$. This issue is of importance for developing single-electron
devices with optical gating based on molecular nanojunctions.

When $\frac{\Omega^{2}\Gamma_{M}^{2}}{4\bar{\mu}^{2}}\gg1,$ the magnitude
$I_{AS}$ is given by \cite{Aku92}%

\[
I_{AS}=\frac{\pi\Gamma_{M}\Omega^{2}}{2\left\vert \bar{\mu}\right\vert
\sqrt{\Gamma_{M}^{2}+\Omega^{2}}},
\]
and we get a simple formula for the charge transferred in the course of slow
passage through the resonance (with respect to both the electron transfer rate
and the reciprocal Rabi frequency)
\begin{equation}
Q=\frac{\pi\Omega^{2}e(\Gamma_{ML,2}-\Gamma_{ML,1})}{2|\bar{\mu}|\sqrt
{\Omega^{2}+\Gamma_{M}^{2}}} \label{eq:Qasympt}%
\end{equation}
Eq.(\ref{eq:Qasympt}) gives $Q=e(\Gamma_{ML,2}-\Gamma_{ML,1})\frac{\pi\Omega
}{2\left\vert \bar{\mu}\right\vert }\gg e$ at least for strong interaction
when $\Omega^{2}\gg\Gamma_{M}^{2}$.

\subsubsection{Slow passage through the resonance and strongly chirped pulses}

Eq.(\ref{eq:Qasympt}) can be obtained directly by integrating
Eq.(\ref{eq:I_Ad1}) with respect to time for $\omega(t)=\omega_{0}-\bar{\mu}t$
and $\Omega=const$. Indeed, integrating Eq.(\ref{eq:I_Ad1}) yields
\begin{align}
Q  &  =\int_{-\infty}^{\infty}I(t)dt=\frac{\Gamma_{M2}+\Gamma_{M1}}%
{4\Gamma_{M1}\Gamma_{M2}}\times\nonumber\\
&  \times\frac{\pi\Omega^{2}e(\Gamma_{ML,2}\Gamma_{MR,1}-\Gamma_{ML,1}%
\Gamma_{MR,2})}{|\bar{\mu}|\sqrt{\Omega^{2}\frac{(\Gamma_{M2}+\Gamma_{M1}%
)^{2}}{4\Gamma_{M1}\Gamma_{M2}}+(\Gamma_{MN}/2)^{2}}} \label{eq:Q}%
\end{align}
In the special case $\Gamma_{M1}=\Gamma_{M2}\equiv\Gamma_{M}$ and
$B_{N}(\varepsilon_{2}-\varepsilon_{1})=0$, Eq.(\ref{eq:Q}) leads to
Eq.(\ref{eq:Qasympt}). As a matter of fact, Eq.(\ref{eq:Q}) extends the case
of slow passage through the resonance beyond the treatment of Ref.\cite{Aku92}.

Eq.(\ref{eq:Q}) can be used for the excitation of a bridging molecule by
Gaussian pulses, Eqs.(\ref{eq:gausspulse}) and (\ref{eq:deltamu}), as well
when the pulses are strongly chirped \cite{Fai00CPL}%

\begin{equation}
2|\Phi^{\prime\prime}(\omega)|\gg\tau_{p0}^{2} \label{eq:strongchirp}%
\end{equation}
For a strongly chirped pulse, one can ascribe to different instants of time
the corresponding frequencies \cite{Fai00CPL}, i.e. different frequency
components of the field are determined via values of the instantaneous pulse
frequency $\omega(t)$ for different instants of time. Then one can integrate
$\int_{-\infty}^{\infty}I(t)dt$ similar to Eq.(\ref{eq:Q}), bearing in mind
that $\Omega(t)=(\frac{d\mathcal{E}_{0}}{\hbar})\exp(-\frac{1}{2}\delta
^{2}t^{2})\approx(\frac{d\mathcal{E}_{0}}{\hbar})\exp(-\frac{\tau_{p0}^{2}%
}{4\Phi^{\prime\prime2}\left(  \omega\right)  }t^{2})$ is a much slower
function of time than $\omega(t)=\omega_{0}-\bar{\mu}t\approx\omega_{0}%
+\frac{1}{\Phi^{\prime\prime}\left(  \omega\right)  }t$. Using
Eqs.(\ref{eq:gausspulse}), (\ref{eq:deltamu}), (\ref{eq:Q}) and
(\ref{eq:strongchirp}), we then get
\begin{align}
Q  &  \approx\frac{\Gamma_{M2}+\Gamma_{M1}}{4\Gamma_{M1}\Gamma_{M2}}%
\times\nonumber\\
&  \times\frac{\sqrt{\pi}(\frac{d}{\hbar})^{2}\mu_{0}cE_{p}\tau_{p0}%
e(\Gamma_{ML,2}\Gamma_{MR,1}-\Gamma_{ML,1}\Gamma_{MR,2})}{\sqrt{(\frac
{d}{\hbar})^{2}\frac{\mu_{0}cE_{p}\tau_{p0}}{\sqrt{\pi}\left\vert \Phi
^{\prime\prime}\left(  \omega\right)  \right\vert }\frac{(\Gamma_{M2}%
+\Gamma_{M1})^{2}}{4\Gamma_{M1}\Gamma_{M2}}+(\Gamma_{MN}/2)^{2}}}
\label{eq:Qstrong_chirp}%
\end{align}
where $\mathcal{E}_{0}^{2}\approx\frac{\mu_{0}cE_{p}\tau_{p0}}{\sqrt{\pi
}\left\vert \Phi^{\prime\prime}\left(  \omega\right)  \right\vert }$, since
the magnitude $\int_{-\infty}^{\infty}\mathcal{E}^{2}(t)dt=2\mu_{0}%
cE_{p}=const$ is conserved in the course of chirping. Here $E_{p}$ is the
pulse energy per unit area, $\mu_{0}$ is the permeability constant, $c$ -
light velocity in vacuum. According to Eq.(\ref{eq:Qstrong_chirp}), in the
case of slow passage through the resonance, $Q\sim\sqrt{\left\vert
\Phi^{\prime\prime}\left(  \omega\right)  \right\vert }$ for strong
interaction when $(\frac{d}{\hbar})^{2}\frac{\mu_{0}cE_{p}\tau_{p0}}{\sqrt
{\pi}\left\vert \Phi^{\prime\prime}\left(  \omega\right)  \right\vert }%
\frac{(\Gamma_{M2}+\Gamma_{M1})^{2}}{4\Gamma_{M1}\Gamma_{M2}}>>(\Gamma
_{MN}/2)^{2}$, and $Q$ tends to a constant value for large $\left\vert
\Phi^{\prime\prime}\left(  \omega\right)  \right\vert $. This elucidates the
behavior observed in our simulations shown in Figs. \ref{fig:chirp_Fi} and
\ref{fig:chirp_Fi_4corr} for moderate and large values of $\left\vert
\Phi^{\prime\prime}\left(  \nu\right)  \right\vert $. In addition,
Eq.(\ref{eq:Qstrong_chirp}) explains why the growth of $Q$ slows down for
larger value of $\left\vert \Phi^{\prime\prime}\left(  \nu\right)  \right\vert
$ if pulse energy increases.

\section{Conclusion}

\label{sec:Conclusion}

In this work a theory for light-induced current by strong optical pulses in
molecular-tunneling junctions have been developed. We have considered a
molecular bridge represented by its highest occupied and lowest unoccupied
levels, HOMO and LUMO, respectively, and have derived a closed set of
equations for electron populations of molecular states, molecular polarization
and molecular excitation (exciton population) when two types of couplings
between the molecule and the metal leads are presented: electron transfer that
gives rise to net current in the biased junction and energy transfer between
the molecule and electron-hole excitations in the leads.

We have used this formalism to analyze a novel control mechanism by which the
charge flow is enhanced by chirped pulses. For linear chirp and when the
energy transfer between the molecule and electron-hole excitations in the
leads is absent, this control model can be reduced to the Landau-Zener
transition to a decaying level, which has an exact analytical solution.

The relaxation parameters in the derived closed set of the equations of motion
do not depend on the exciting electromagnetic field. This is true if the Rabi
frequency $\Omega$ is much smaller than the bath correlation frequency,
$\omega_{c}$. If molecular states $\varepsilon_{m}$ are far from the Fermi
levels of both leads, $\omega_{c}$ is determined by the frequency interval for
the system-bath interaction matrix elements $V_{km}^{(MK)}$ and $V_{kk^{\prime
}}^{(NK)}$ and the density of states of metal leads. The last can be evaluated
as 1-10 eV \cite{Schreiber06}. As a matter of fact, the approximation of
constant relaxation parameters, which do not depend on exciting
electromagnetic radiation, is consistent to the RWA used in our theory.

The situation is different if we assume that the molecular level position is
pinned to the Fermi energy of a lead, that may lead to highly nonlinear
current voltage dependence \cite{Nitzan06JCP}. In this case $\omega_{c}$ is
determined also by the frequency interval at which $f_{K}(\varepsilon)$ is
essentially changed that is $\sim k_{B}T/\hbar$ (see Eqs.(\ref{eq:WMKm}) and
(\ref{eq:BNK})). In the last case $\Omega$ can be of the same order of
magnitude as $\omega_{c}$ in the RWA, and the dependence of the relaxation
parameters on exciting electromagnetic field \cite{Schreiber06} must be
included into the theory.

To end this discussion we note that in this work we have
investigated a model process driven by light absorption in a
molecular bridge connecting metal leads. As already discussed, the
geometry considered is potentially advantageous because of the
possible local field enhancement due to plasmon excitation in the
leads. It should be emphasized however that other processes, not
considered in this work, may play important roles in nanojunction
response to incident light. First, direct electron-hole excitations
of the metal leads \cite{Petek97,Ueba07} may affect response in an
adsorbed molecule that goes beyond the local field enhancement
associated with plasmon excitation. Secondly, experimental
realization of strong local excitations in nanojunctions requires
careful consideration of heating and heat dissipation and conduction
\cite{Nitzan07Jphys}. Heating may be kept under control by driving
the junction using a sequence of well separated optical pulses, as
envisioned in the proposed experiment, but it should be kept in mind
that a more detailed consideration of this issue may be needed.

\textbf{Acknowledgement}

This work was supported by the Israeli Science Foundation (BF and AN), the
German-Israeli Fund (AN) and the US-Israel BSF (AN).

\appendix

\section{}

Calculate the steady-state current in the absence of the radiative and
nonradiative energy transfer couplings, $\hat{V}_{P}$ and $\hat{V}_{N}$. The
corresponding solution of Eq.(\ref{eq:n_m1a}) is as follows: $n_{m}%
=W_{Mm}/\Gamma_{Mm}$. Substituting it into Eq.(\ref{eq:I_1}) and using
Eq.(\ref{eq:WMKm1}), we get%
\begin{equation}
I=e\sum_{m=1,2}\frac{\Gamma_{ML,m}\Gamma_{MR,m}}{\Gamma_{Mm}}[f_{R}%
(\varepsilon_{m})-f_{L}(\varepsilon_{m})] \label{eq:I_App}%
\end{equation}
The last formula can be written%
\begin{widetext}
\begin{eqnarray}
I=e\sum_{m=1,2}\frac{\Gamma_{ML,m}\Gamma_{MR,m}}{\Gamma_{Mm}}\int
d\varepsilon\lbrack f_{R}(\varepsilon)-f_{L}(\varepsilon)]\delta
(\varepsilon-\varepsilon_{m})=\nonumber\\
=\frac{e}{2\pi\hbar}\sum_{m=1,2}\lim_{\Gamma_{Mm}/2\rightarrow0}
\Gamma_{ML,m}\Gamma_{MR,m}\int
d\varepsilon\frac{f_{R}(\varepsilon)-f_{L}(\varepsilon
)}{(\varepsilon-\varepsilon_{m})^{2}\hbar^{-2}+[\Gamma_{Mm}/2]^{2}}
\label{eq:I_App1}
\end{eqnarray}
\end{widetext}
using the well known representation for $\delta(x)$%
\begin{equation}
\delta(x)=\frac{1}{\pi}\lim_{\sigma\rightarrow0}\frac{\sigma}{x^{2}+\sigma
^{2}} \label{eq:delta}%
\end{equation}
The limit $\lim_{\Gamma_{Mm}/2\rightarrow0}$ on the right-hand side of
Eq.(\ref{eq:I_App1}) is consistent with the Markovian approximation in the
sense that relaxation parameters $\Gamma_{Mm}/2$ are small in comparison to
the bath correlation frequency, $\omega_{c}$.

The term $\lim_{\Gamma_{Mm}/2\rightarrow0}\frac{1}{(\varepsilon-\varepsilon
_{m})^{2}\hbar^{-2}+[\Gamma_{Mm}/2]^{2}}$ on the right-hand side of
Eq.(\ref{eq:I_App1}) can be written
\begin{equation}
\lim_{\Gamma_{Mm}/2\rightarrow0}\frac{1}{(\varepsilon-\varepsilon_{m}%
)^{2}\hbar^{-2}+[\Gamma_{Mm}/2]^{2}}=G_{mm}^{r}(\varepsilon)G_{mm}%
^{a}(\varepsilon) \label{eq:Green}%
\end{equation}
where $G_{mm}^{r}(\varepsilon)$ and $G_{mm}^{a}(\varepsilon)$ are the retarded
and advanced Green's functions, respectively, \cite{Nitzan06JCP} in the
Markovian approximation. Substitution of Eq.(\ref{eq:Green}) into
Eq.(\ref{eq:I_App1}) leads to the well known Landauer formula for the current
\cite{Hau96}.

\section{}

Let us show that $n_{1}=1-n_{2}$ when $\Gamma_{M1}=\Gamma_{M2}\equiv\Gamma
_{M}$. Summing Eqs.(\ref{eq:n_1}) and (\ref{eq:n_2}) at given condition, we have%

\begin{equation}
\frac{dy}{dt}=-\Gamma_{M}y
\end{equation}
where we denoted $y=n_{1}+n_{2}-1$. The solution of the last equation is as follows%

\begin{equation}
y(t)=y(0)\exp(-\Gamma_{M}t)
\end{equation}
where $y(0)=0$. This gives $n_{1}(t)+n_{2}(t)=1$ even in the presence of
energy transfer when $B_{N}(\varepsilon_{2}-\varepsilon_{1})\neq0$.

%\bibliographystyle{osajnl}
%\bibliography{FeinbergArt.bib}

\end{document}